\documentclass[twocolumn,superscriptaddress,10pt,aps,prb]{revtex4-2}
\usepackage{epsfig,amsopn}
\usepackage{graphicx}
\usepackage{epstopdf}
\usepackage{braket}
\usepackage{amsmath,amssymb}
\usepackage{amsthm}
\usepackage{enumerate}
\usepackage{xcolor}
\usepackage{fontawesome5}
\usepackage{mathtools}
\usepackage{titlesec}
\usepackage{enumitem}

\usepackage[colorlinks=true,linkcolor=blue,allcolors=blue,
filecolor=black,citecolor = magenta,urlcolor=magenta]{hyperref}
\definecolor{orcidlogocol}{rgb}{0.65, 0.807, 0.223}
\newcommand{\orcid}[1]{$$\href{https://orcid.org/#1}{\textcolor{orcidlogocol}{\faOrcid}}}
\usepackage[defaultcolor=blue]{changes}
\usepackage{todonotes}

\titlespacing*{\section}
{0pt}{*3}{*1.5}   % {left}{before}{after}

\titlespacing*{\subsection}
{0pt}{*2}{*1}

\newcommand\vect{\boldsymbol}

\newcommand\beq{\begin{equation}}
	\newcommand\eeq{\end{equation}}
\newcommand\bes{\begin{subequations}}
	\newcommand\ees{\end{subequations}}
\newcommand\bea{\begin{eqnarray}}
	\newcommand\eea{\end{eqnarray}}

\newcommand\la{\langle}
\newcommand\ra{\rangle}

\newcommand{\RNum}[1]{\uppercase\expandafter{\romannumeral #1\relax}}

\begin{document}
	\title{Quantized Transport in Floquet Topological Insulators}  
	\author{Rekha Kumari}
	\email{rekha.kumari@icts.res.in}
	\email{rekha.kumari.e6@tohoku.ac.jp}
    \affiliation{International Centre for Theoretical Sciences, Tata Institute of Fundamental Research, Bangalore 560089, India}
    \affiliation{Advanced Institute for Materials Research (WPI-AIMR), Tohoku University, Sendai 980-8577, Japan}
	\author{Manas Kulkarni}
    \email{manas.kulkarni@icts.res.in}
	\affiliation{International Centre for Theoretical Sciences, Tata Institute of Fundamental Research, Bangalore 560089, India}
	\author{Abhishek Dhar}
    \email{abhishek.dhar@icts.res.in}
	\affiliation{International Centre for Theoretical Sciences, Tata Institute of Fundamental Research,
		Bangalore 560089, India}
	
	\date\today
    \begin{abstract}
    {We study quantum transport in a periodically driven (Floquet) topological system coupled to static fermionic reservoirs. Using the Floquet nonequilibrium Green’s-function (NEGF) formalism we show, from exact numerics for  a strip geometry, that the two-terminal (longitudinal) conductance is quantized as $|W_{\varepsilon}|\,e^2/h$, while the Hall (transverse) conductance is quantized as $W_{\varepsilon}\,e^2/h$, where $W_{\varepsilon}$ is the Floquet winding invariant associated with the quasienergy gap at $\varepsilon = 0$ or $\varepsilon = \Omega/2$. 
    Quantization is achieved only after summing over the contribution of all Floquet sidebands. We provide an analytic understanding of this Floquet conductance sum rule, by considering the Hall conductance in the weak coupling limit. In that limit, we show that the Floquet Hall conductance gets contributions from the Floquet sidebands, which includes the signs of the velocities of the edge modes. Their sum yields exact quantization, as predicted by the Floquet sum rule. We find that in a wide range of parameter regime, the convergence is fast, making observation of the sum rule and Floquet winding numbers accessible to experiments.}
	\end{abstract}

	\maketitle
	\section{Introduction}
	Periodic driving offers a versatile route to engineering quantum phases that have no static counterparts~\cite{Bukov2015,Goldman2014,Eckardt2017,Harper2020,Oka2019,FTI_Moessner,HOFTP_Vega,FTP_Vishwanath,Phonon-induced-FTP-Swati,FTP-Foa-Torres,FTPT_Kundu_Seradjeh,FTPT-cold-atoms,Bukov04032015,Valley_filter_rekha,Kumari2024,6wnf-b5g8,roy2026floquetgenerationhybridordertopology,PhysRevB.98.054203,qvdz-qwf8}. In two-dimensional systems, Floquet driving can generate chiral edge modes at both zero and $\Omega/2$ quasienergies, enabling ``Floquet anomalous topological phases" in which all bulk Chern numbers vanish while protected edge transport persists~\cite{Rudner2013,Unal2019PRR,Nathan2015,Titum2016,Quelle_2017,Rudner2020,Disorder-FTI-Refael}. These phases are characterized not by the Chern numbers but by the Floquet winding invariants defined in combined momentum--time space, which count the net number of chiral modes crossing the quasienergy gaps at $0$ and $\Omega/2$~\cite{Rudner2013,Nathan2015}. Experimental realizations of such anomalous Floquet phases featuring chiral edge modes have been demonstrated in driven photonic lattices~\cite{Mukherjee2017,topological_photonic}. Related phases were also shown to exhibit quantized nonadiabatic charge pumping governed by winding invariants in disordered systems, known as anomalous Floquet--Anderson insulators (AFAI)~\cite{Titum2016}. While the bulk and boundary properties of these Floquet phases are well understood~\cite{Rudner2013,Hopf_T3}, their signatures in transport experiments remain less clear.
	
	In static topological systems, the two-terminal and Hall conductances exhibit quantization when edge states equilibrate with fermionic reservoirs, providing a direct probe of bulk topology~\cite{Klitzing1980,Thouless1982,Buttiker1988,Hasan2010,Bernevig2006,Koenig2007,Mera_n_Ozawa,BuroMera_work_topoquant,sinha2025proximityeffectstopologicalinvariant,Majeed_Bhat_2025}. In contrast, in periodically driven systems, numerical studies have shown that quantized conductance can be restored by applying Floquet sum rules, first introduced in the context of two-terminal transport signatures of Floquet Majorana fermions~\cite{Seradjeh_PRL}, 
    where it was demonstrated both analytically and numerically, that summing the conductance over bias values separated by integer multiples of drive frequency yields a quantized result.
    Subsequently, it has also been shown that summing over all photon-assisted transmission channels leads to quantized two-terminal longitudinal conductance in Floquet topological insulators~\cite{Foa_Torres_2014,Esin2018,Dehghani2015,Farrell2016,Yap2017,Bajpai2020}. However, the microscopic origin of these sum rules in open transport setups, and their direct connection to Floquet winding invariants~\cite{Rudner2013}, remain poorly understood. Recent work has shown that quantized transport in Floquet systems can also be restored by engineering narrow-band energy filters that suppress photon-assisted processes~\cite{Netanel_2024_Quanitized_transport}, demonstrating the recovery of quantization without invoking additional Floquet sum rules.
	
	Recently, it has been shown that when a periodically driven system is coupled to static fermionic reservoirs, the system reaches a steady state in the weak-coupling limit~\cite{Kohler2005,Iadecola2015,Kumari2024,Matsyshyn2023,Valley_filter_rekha,b3pw-my97}, in which the spectral weight of Floquet--Bloch states is distributed across multiple photon sidebands. In this Floquet steady state, the occupations of floquet quasi energy bands deviate significantly from equilibrium distributions due to the periodic drive. In the context of the Josephson effect of Floquet Majorana fermions, it was shown that in the steady state achieved in the weak-coupling limit, the $4\pi$-periodic Josephson effect--the hallmark of Majorana fermions--can be understood in terms of quantized occupation differences summed over quasienergy sidebands~\cite{Kumari2024}. As a result, transport quantization is generally not recovered within any single Floquet zone. Consequently, the Josephson current–phase relation for Floquet Majorana fermions closely resembles that of the static case when expressed using Floquet sum rules.
	
    Similarly, in the context of the Floquet topological insulators, it remains unclear how the steady state of an open Floquet system encodes information about Floquet winding invariants, particularly in anomalous phases with vanishing Chern numbers. A unified microscopic framework directly connecting steady-state occupations, winding invariants, and measurable transport signatures is therefore still lacking. To address this issue, we show that the Floquet conductance sum rule, widely used to recover quantized transport signatures of Floquet topological insulators, emerges naturally from the redistribution of the spectral density of chiral Floquet edge modes in the steady state. Moreover, we compute the Hall conductance, which tracks not only the absolute value but also the sign of the winding invariant. 
	
	\begin{figure*}[t]
		\centering
		\includegraphics[width=1.\linewidth]{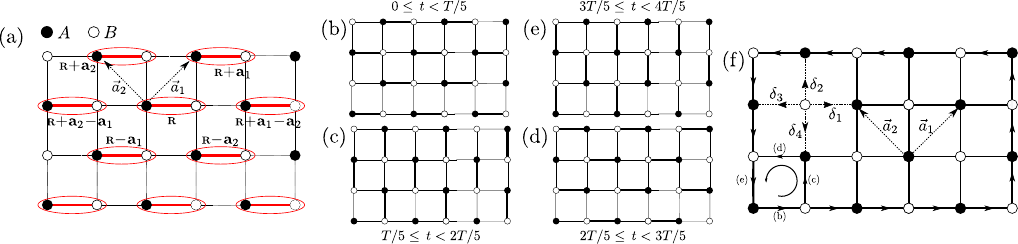}
		\caption{(a) Bipartite lattice generated by diagonal primitive vectors $\vect a_1$ and $\vect a_2$, unit cell structure is highlighted by red circles, each unit cell contains two sub-lattices marked by $A$ and $B$. (b–e) Five-step driving protocol. Hopping with amplitude $J$ is sequentially activated along four bond directions (highlighted links). (f) Effective stroboscopic dynamics over one driving period $T$. The sequential displacements (a-d), generate chiral hopping along the bonds, resulting in a net chiral circulation around each plaquette and an emergent unidirectional Floquet dynamics.}
		\label{fig:FIG1_SYS}
	\end{figure*}
	In this work, we develop a unified theoretical framework that links Floquet topology and quantized transport to the steady state of driven open quantum systems. The main results of this work are summarized as follows:
	
	\begin{enumerate}
		\item We first consider the strip geometry and using the Floquet nonequilibrium Green’s function (NEGF) formalism~\cite{Kohler2005}, we demonstrate quantization of the two terminal conductance using the Floquet sum rules. We show that, in certain parameter regimes, the sum rule is even satisfied on including a small number of Floquet sidebands.

        \item We show, again for the strip geometry, that the spatially resolved bond currents (summed over the Floquet sidebands) can be used to compute, in addition to the two-terminal conductance, also the Hall conductance which encodes both the magnitude and the sign of the Floquet winding invariants.
        
		\item We analyze the dependence of both the two terminal and Hall conductance quantization on system size and system--reservoir coupling strength, and show that quantization emerges in the weak-coupling limit when the transverse system size is sufficiently large. This result is valid both for the strip and cylindrical geometries.
	
		\item In the cylindrical geometry and in the weak coupling limit, we provide 
         a microscopic interpretation of the Floquet sum rule and 
        an analytic proof of the Hall conductance quantization. We show that summing over Floquet sidebands in the weak-coupling steady state reconstructs the full spectral weight of chiral Floquet edge modes distributed over Floquet replicas, thereby yielding an integer Hall response set by the winding invariants.
	\end{enumerate}
	
	The paper is organized as follows. In Sec.~\ref{sec:Floquet topological phases}, we introduce the driven lattice model and its Floquet topological phases. In Sec.~\ref{sec:NEGF_formalism_for_Floquet_systems}, we develop the Floquet–NEGF formalism and present expressions for the two-terminal (longitudinal) and Hall conductances using spatial bond currents on the lattice. In Sec.~\ref{sec:Numerical results}, we present numerical results for the two-terminal and bond-resolved conductances in a strip geometry, and subsequently analyze the same quantities in a cylindrical setup. In the weak coupling limit, we demonstrate conductance quantization and its relation to Floquet winding invariants. We also provide an analytical derivation of the Hall conductance and present the microscopic origin of the Floquet sum rule. Sec.~\ref{sec:Summary and Outlook} concludes with implications for experiments.
    
    For convenience, the notation used throughout the manuscript is summarized in Appendix~\ref{sec:Glossary of symbols}, and additional technical details are provided in the appendices. Throughout this paper we set $e=1$ and $\hbar=1$.
	
	\section{Model, lattice geometries and Floquet topological invariants} 
	\label{sec:Floquet topological phases}
    To study transport in Floquet topological insulators (FTIs), we consider the minimal two-band lattice model introduced in Ref.~\cite{Rudner2013}. The system is defined on a bipartite square lattice as shown in Fig.~\ref{fig:FIG1_SYS}(a), with a two site unit cell with Bravais lattice vectors $\mathbf a_1 = (1,1)$ and $\mathbf a_2 = (-1,1)$ where we set lattice spacing to unity. Each unit cell, with  position,
    \begin{equation}
        \mathbf R = m\mathbf a_1 + n\mathbf a_2,
    \end{equation}
    contains two sublattice sites, denoted by $A$ and $B$, as illustrated in Fig.~\ref{fig:FIG1_SYS}(a). Here $m=1,\ldots,n_x$ and $n=1,\ldots,n_y$ (where $n_y$ is chosen to be even), so that the total number of unit cells is $n_x n_y$. The underlying lattice (that includes both sublattice sites) can alternatively be labeled by integer coordinates $\mathbf r=(x,y)$ with $x=1,\ldots,2n_x$ and $y=1,\ldots,n_y$. The lattice sites satisfying $x+y$ even (odd) belong to sublattice $A$ ($B$). The total number of lattice sites is therefore $2n_x\times n_y$.

	Under the five-step driving protocol introduced in Ref.~\cite{Rudner2013}, the lattice Hamiltonian over five equal time intervals of duration $T/5$, can be written as
	\begin{align}
		H(t)=H_j(t), \qquad t \in \left[\frac{(j-1)T}{5},\, \frac{jT}{5}\right].\label{eq:h5step_strip}
	\end{align}
	Here $j=1,\ldots,5$. Throughout the drive, the staggered sub-lattice potential $M$ is present. During the first four steps of the drive, the hopping amplitude $J$ is sequentially activated along different bond orientation, as shown in Fig.~\ref{fig:FIG1_SYS}(b-e), respectively. In the fifth step 
	all the hoppings are switched off, thereby leaving only the sub-lattice potential on. Repeating the five-step protocol explicitly breaks time-reversal symmetry and produces a net chiral motion over one driving period, leading to Floquet chiral edge modes, as shown in Fig.~\ref{fig:FIG1_SYS}(f). The lattice Hamiltonian in the first four hopping steps of the drive can then be compactly written using the two-component basis vectors 
    \begin{equation}
    \psi(\mathbf R)=
    \begin{pmatrix}
	c_A(\mathbf R) \\
	c_B(\mathbf R)
    \end{pmatrix}\,,
    \end{equation} as
	\begin{align}
		H_j &= -J \sum_{\mathbf r}
		\left[
		\psi^\dagger(\mathbf R)\sigma_+\psi(\mathbf R+\mathbf d_j)
		+ \mathrm{h.c.}
		\right]
		+ H_5,\\
		H_5 &= M \sum_{\mathbf R}
		\psi^\dagger(\mathbf R)\sigma_z\psi(\mathbf R)\,,
	\end{align}
    Here $j=1,\ldots,4$ and $d_j$ are the step-dependent translation vectors, given by
	\begin{equation}
		\mathbf d_1 = 0, \qquad
		\mathbf d_2 = \mathbf a_2, \qquad
		\mathbf d_3 = \mathbf a_2 - \mathbf a_1, \qquad
		\mathbf d_4 = -\mathbf a_1 .
	\end{equation}
	Note that $\mathbf d_1=0$ corresponds to hopping within the same Bravais unit cell, while the remaining $\mathbf d_j$ generate the other inter-cell hoppings.
	
    We now perform a Fourier transform with respect to the Bravais lattice vectors $\mathbf a_1,\mathbf a_2$, 
    $\psi(\mathbf R)
	=
	\frac{1}{\sqrt{N}}
	\sum_{\mathbf k}
	e^{i\mathbf k\cdot\mathbf R}
	\psi(\mathbf k)$,
    where $N=n_x n_y$. The crystal momentum takes values
	\begin{align}
    \mathbf k = \frac{p}{n_x}\mathbf b_1+\frac{q}{n_y}\mathbf b_2,
    \end{align}
	with $p=0,\ldots,n_x-1$ and $q=0,\ldots,n_y-1$, where the reciprocal lattice vectors satisfy
	\(
	\mathbf a_i\cdot\mathbf b_j = 2\pi \delta_{ij}.
	\)
	For $\mathbf a_1=(1,1)$ and $\mathbf a_2=(-1,1)$, the reciprocal vectors are $\mathbf b_1=\pi(1,1)$ and $\mathbf b_2=\pi(-1,1)$. Using this transformation, the Hamiltonian can be written in momentum space as
	\begin{equation}
		H_j(t)=\sum_{\mathbf k}
		\psi^\dagger(\mathbf k)
		h_j(\mathbf k,t)
		\psi(\mathbf k),\label{eq:h5_kxkyt}
	\end{equation}
	with the step-dependent Bloch Hamiltonians
	\begin{align}
		h_1(\mathbf k) &= -J\,e^{i\mathbf k\cdot\boldsymbol\delta_1}\sigma_+ + \mathrm{h.c.} + M\sigma_z, \\
		h_2(\mathbf k) &= -J\,e^{i\mathbf k\cdot\boldsymbol\delta_2}\sigma_+ + \mathrm{h.c.} + M\sigma_z, \\
		h_3(\mathbf k) &= -J\,e^{-i\mathbf k\cdot\boldsymbol\delta_1}\sigma_+ + \mathrm{h.c.} + M\sigma_z, \\
		h_4(\mathbf k) &= -J\,e^{-i\mathbf k\cdot\boldsymbol\delta_2}\sigma_+ + \mathrm{h.c.} + M\sigma_z ,\\
		h_5(\mathbf k) &= M\sigma_z,
	\end{align}
    where $\boldsymbol\delta_1=(1,0)$, $\boldsymbol\delta_2=(0,1)$, $\boldsymbol\delta_3=(-1,0)$ and $\boldsymbol\delta_4=(0,-1)$, which satisfy
	\(
	\mathbf a_1 = \boldsymbol\delta_1 + \boldsymbol\delta_2
	\)
	and
	\(
	\mathbf a_2 = -\boldsymbol\delta_1 + \boldsymbol\delta_2.
	\)
	Floquet theory provides a complete set of eigenstates for periodically driven systems satisfying $h(\mathbf{k},t)=h(\mathbf{k},t+T)$. The solutions of the time-dependent Schr\"odinger equation given by
	\begin{align}
		i\partial_t |\psi_{\alpha}(\mathbf{k},t)\rangle=h(\mathbf{k},t)|\psi_{\alpha}(\mathbf{k},t)\rangle\label{eq:tdshe}
	\end{align}
	can then be written as Floquet states:
	\begin{equation}
		|\psi_{\alpha}(\mathbf{k},t)\rangle
		=
		e^{-i\epsilon_{\alpha}(\mathbf{k})t}
		|\phi_{\alpha}(\mathbf{k},t)\rangle,
		\label{eq:ualpha}
	\end{equation}
	where $|\phi_{\alpha}(\mathbf{k},t)\rangle$ are time-periodic functions known as Floquet modes, i.e.,
	$|\phi_{\alpha}(\mathbf{k},t)\rangle = |\phi_{\alpha}(\mathbf{k},t+T)\rangle$, and
	$\epsilon_{\alpha}(\mathbf{k})$ are known as quasienergies. Substituting
	Eq.~\eqref{eq:ualpha} into the Schr\"{o}dinger equation~\eqref{eq:tdshe} yields
	\begin{equation}\label{eq:eigen_value_eq_Floquet_mods}
		\left[ h(\mathbf{k},t) - i \partial_t \right]
		|\phi_{\alpha}(\mathbf{k},t)\rangle
		= \epsilon_{\alpha}(\mathbf{k}) |\phi_{\alpha}(\mathbf{k},t)\rangle,
	\end{equation}
	an eigenvalue equation for Floquet modes $|\phi_{\alpha}(\mathbf{k},t)\rangle$.

	Due to the time-periodic nature of both the Hamiltonian and the Floquet eigenmodes, the Floquet eigenvalue equation can be transformed into Fourier space. We introduce the following Fourier expansions for the Bloch Hamiltonian and the Floquet modes,
	\begin{align}
    \label{eq:floqfour}
		h(\mathbf{k},t) &= \sum_{n=-\infty}^{\infty} e^{-i n \Omega t}\, h^{(n)}(\mathbf{k}), \\
		|\phi_{\alpha}(\mathbf{k},t)\rangle &= \sum_{m=-\infty}^{\infty} e^{-i m \Omega t}\,
		|\phi^{(m)}_{\alpha}(\mathbf{k})\rangle,
	\end{align}
	where $\Omega = 2\pi/T$ is the driving frequency. Substituting these expansions into the Floquet eigenvalue equation~\eqref{eq:eigen_value_eq_Floquet_mods} yields a time-independent eigenvalue problem in the enlarged Floquet--Sambe space (see Appendix~\ref{app:EZF} for details). Diagonalization of the Floquet Hamiltonian (with a suitable numerical cut-off for $n,m$) yields the quasienergy spectrum and the associated Floquet modes.
	
	Alternatively, Floquet eigenstates can also be obtained from the one-period time-evolution operator,
	\begin{equation}
		U_{\mathbf{k}}(T)
		=
		\mathcal{T}
		\exp\!\left(
		-i\int_{0}^{T}
		h(\mathbf{k},t')\,dt'
		\right),
	\end{equation}
	The eigenvalues of the one-period time-evolution operator $e^{-i\epsilon_{\alpha}(\mathbf{k})T}$, define the quasienergy bands $\epsilon_{\alpha}(\mathbf{k})$. The Chern number associated with a Floquet quasienergy band can then be defined analogously to static systems as
	\begin{equation}
		C_\alpha
		=
		\frac{i}{2\pi}
		\int_{\mathrm{BZ}} d^2k\;
		\left(
		\langle \partial_{k_x}\xi_\alpha | \partial_{k_y}\xi_\alpha \rangle
		-
		\langle \partial_{k_y}\xi_\alpha | \partial_{k_x}\xi_\alpha \rangle
		\right),
		\label{eq:Chern-invariant}
	\end{equation}
	where $\xi_\alpha \equiv \xi_\alpha(\mathbf{k})$ is the eigenstate associated with the $\epsilon_{\alpha}(\mathbf k)$ quasienergy band, i.e. $U_{\mathbf{k}}(T)|\xi_\alpha \rangle =e^{-i\epsilon_{\alpha}(\mathbf{k})T}|\xi_\alpha \rangle$. The Chern number $C_\alpha$ characterizes the topology of $\alpha$-th Floquet quasienergy bands~\cite{SeshadriSen2022}. 
	Unlike in the static case, in periodically driven systems, such as the one described in Eq.~\eqref{eq:h5_kxkyt}, the Chern number does not fully characterize the presence of Floquet edge modes.  Additional topological invariants of the time-dependent Floquet systems are given by the integer-valued winding numbers
    \begin{eqnarray}
    W[U_{\epsilon}]&=&\frac{1}{8\pi^2}\int d^2k\,dt\;\nonumber\\
   && \times \mathrm{Tr}\!\left(
    U_{\epsilon}^{-1}\partial_t U_{\epsilon}
    \bigl[
    U_{\epsilon}^{-1}\partial_{k_x}U_{\epsilon},
    U_{\epsilon}^{-1}\partial_{k_y}U_{\epsilon}
    \bigr]
    \right),
    \label{eq:winding_def_main}
    \end{eqnarray}
    which is invariant under smooth deformations that preserve quasienergy gaps at $\epsilon=0$ and $\Omega/2$. The details of the operator $U_{\epsilon}$ and winding invariant calculations are provided in Appendix~\ref{sec: Floquet winding invariants}.

	\begin{figure}[t]
		\centering
        \includegraphics[width=0.75\linewidth]{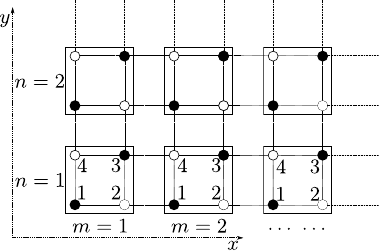}
        \caption{Schematic diagram of the cylindrical setup. The system is periodic along the $y$ direction, and finite along the $x$ direction. We indicate the four-component basis using square unit cells. The total number of cells along the $x$ direction (indicated by $m$) are $n_x$ and the total number of cells along the $y$ direction (indicated by $n$) are $n_y/2$.}
        \label{fig:4_comp_cyl}
    \end{figure}

    In Floquet systems, the explicit time dependence can lead to anomalous topological phases in which the Chern number vanishes, yet the system hosts topological edge states at quasienergies $0$ and $\Omega/2$. To illustrate this, we consider the system on a cylindrical geometry by imposing periodic boundary conditions along the $y$ direction while keeping the $x$ direction open.
	
    To analyze the system on a cylindrical geometry, it is convenient to consider a four lattice site basis as shown in Fig.~\ref{fig:4_comp_cyl}. The unit cell locations are now given by 
    \begin{equation}    {\bf R}=m (2,0)+n(0,2),\quad \text{for cylindrical geometry}
    \end{equation}
    and the $4-$component wavefunction on each unit cell is
    given by
    \begin{align}
    \Psi^\dagger(m,n) =
	\left(
	c^\dagger_{1},
	c^\dagger_{2},
	c^\dagger_{3},
	c^\dagger_{4}
	\right)_{m,n},
    \end{align}
    where $m=1\ldots n_x$ and $n=1\ldots n_y/2$. We perform a partial Fourier transform along the $y$ direction,
	$\Psi(R) = \frac{1}{\sqrt{n_y/2}} \sum_{k_y} e^{i k_y y} \Psi(m,k_y),$
	where $k_y =4\pi s/n_y$ (here $s$ varies from $1$ to $n_y/2$), labels the conserved crystal momentum.

	\begin{figure}[t]
		\centering
		\includegraphics[width=0.85\linewidth]{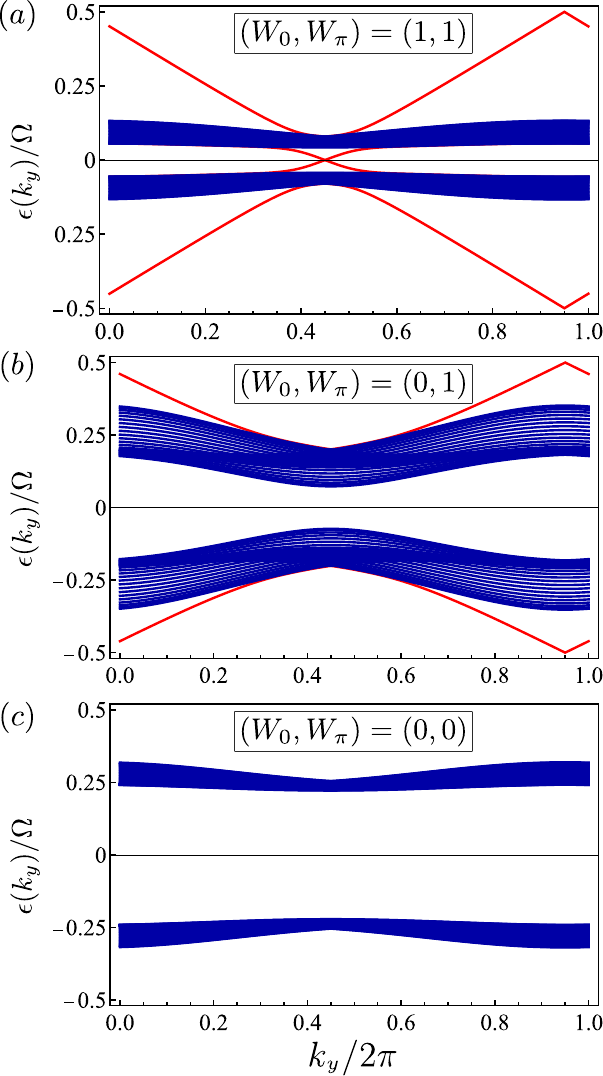}
		\caption{Quasienergy spectrum (in units of $\Omega$) of the Hamiltonian in Eq.~\eqref{eq:H5_cylinder} in the cylindrical geometry, for (a) $J=1.25\Omega$, (b) $J=0.75\Omega$, and  (c) $J=0.25\Omega$, with $M=0.25\Omega$ and $n_x=25$. Chiral edge modes appear in the quasienergy gaps at $0$ and $\Omega/2$ as the ratio $J/\Omega$ is varied.}
		\label{fig:FIG2_QES}
	\end{figure}
    In this four component basis, the Hamiltonian in mixed representation reads
    \begin{equation}
    H_j(t)=\sum_{m,m',k_y}
    \Psi^\dagger(m,k_y)\,
    h_{j,mm'}(k_y)\,
    \Psi(m',k_y),
    \label{eq:h5_kxkyt_cyl}
    \end{equation}
    where the step-dependent Bloch Hamiltonians are given by
    \begin{align}
    h_{1,mm'}(k_y) &= -J\,\sigma\otimes\sigma_-  \,\delta_{m',m+1}
    + h_{5,mm'}(k_y), \\\nonumber
    h_{2,mm'}(k_y) &= -J (\sigma\otimes\sigma_x)+ (\bar{\sigma}\otimes(\sigma_x\cos k_y+\sigma_y\sin k_y))
    \\
    &~~~~~\delta_{m,m'}+ h_{5,mm'}(k_y), \\
    h_{3,mm'}(k_y) &= -J\,\bar{\sigma} \otimes \sigma_+ \,\delta_{m',m+1}
    + h_{5,mm'}(k_y), \\\nonumber
    h_{4,mm'}(k_y) &= -J (\bar{\sigma}\otimes\sigma_x)+ (\sigma\otimes(\sigma_x\cos k_y-\sigma_y\sin k_y))
    \\
    &~~~~~\delta_{m,m'}+ h_{5,mm'}(k_y),\\
    h_{5,mm'}(k_y) &= M\,\mathbb{I}\otimes\sigma_z\,\delta_{m,m'}.
    \end{align}
    Here we define $\sigma \coloneqq \tfrac{1}{2}(\mathbb{I} +\sigma_z)$ and $\bar{\sigma} \coloneqq \tfrac{1}{2}(\mathbb{I} -\sigma_z)$.
	The full Hamiltonian for each momentum sector is therefore
	\begin{align}
	H(t)=H_j(t), \qquad t \in \left[\frac{(j-1)\,T}{5},\, \frac{j\,T}{5}\right].\label{eq:H5_cylinder}
	\end{align}
	Here $j=1,\ldots,5$. The Floquet quasienergy spectrum for the cylindrical system given in Eq.~\eqref{eq:H5_cylinder} is shown in Fig.~\ref{fig:FIG2_QES} for several values of the ratio $J/\Omega$. Panel~(a) shows chiral edge modes at both $0$ and $\Omega/2$, characterized by $(W_0,W_\pi)=(1,1)$ but $C_+=0$, which is the hallmark of an anomalous Floquet topological insulator phase. Panel~(b) displays a single chiral edge mode crossing the quasienergy gap at $\Omega/2$, yielding $(W_0,W_\pi)=(0,1)$ and $C_+=1$. Fig.~\ref{fig:FIG2_QES}(c) corresponds to a topologically trivial phase with winding invariants $(W_0,W_\pi)=(0,0)$, and Chern number $C_+=0$. In the following section, we analyze how these distinct Floquet topological phases manifest in two-terminal transport using the Floquet nonequilibrium Green’s-function formalism.

	\begin{figure}[ht!]
		\centering
		\includegraphics[width=1.\linewidth]{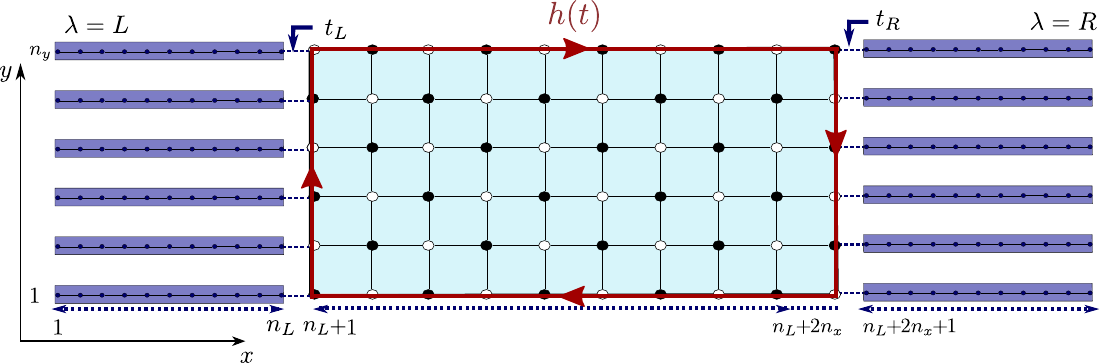}
		\caption{
			Schematic of the two-terminal transport setup in a strip geometry. The central region consists of a periodically driven 2d lattice of size $2n_x\times n_y$. The driven system is coupled to two semi-infinite, non-driven fermionic reservoirs labeled $L$ and $R$, which are attached to the left and right edges of the strip at $\mathbf r=(n_L+1,j)$ and $\mathbf r=(n_L+2n_x,j)$, respectively, for $j=1,\dots,n_y$.
		}
		\label{fig:FIG-SETUP}
	\end{figure}
	\section{NEGF formalism for Floquet systems}\label{sec:NEGF_formalism_for_Floquet_systems}
	To study quantized transport in the Floquet topological system introduced in Eq.~\eqref{eq:h5step_strip}, we consider a two-terminal transport setup in a strip geometry (no periodicity in either direction), as illustrated in Fig.~\ref{fig:FIG-SETUP}. The central region of the figure consists of a periodically driven system  with the total number of sites $2n_x\times n_y$. The basis is given by $c^{\dagger}(r)$, where parity of $r$ defines the sub-lattices. The $x$ and $y$ coordinates denote the longitudinal and transverse directions of the strip, respectively.
	
	The driven central system is coupled to two macroscopic, non-driven fermionic reservoirs labeled by $\lambda=L,R$, as shown in Fig.~\ref{fig:FIG-SETUP}. Each reservoir is modeled as an ideal Fermi reservoir characterized by an equilibrium Fermi--Dirac distribution with chemical potential $\mu_\lambda$ and zero temperature. The reservoirs are attached along the longitudinal edges of the strip: the left reservoir ($L$) couples uniformly to all sites $\mathbf r=(n_L+1,j)$, while the right reservoir ($R$) couples to sites $\mathbf r=(n_L+2n_x,j)$, with $j=1,\ldots,n_y$. The tight-binding Hamiltonian of the full setup is given by
	\begin{equation}
	\mathcal{H}(t)
	=
	H(t)
	+
	\sum_{\lambda=L,R}
	\left(
	H_\lambda + H_{S\lambda}
	\right),
	\label{eq:H5_sys_2d_junction}
	\end{equation}
	where $H(t)$ is the time-dependent Hamiltonian of the driven central region defined in Eq.~\eqref{eq:h5step_strip}, $H_\lambda$ describes the tight binding Hamiltonian of reservoir $\lambda$, and $H_{S\lambda}$ accounts for tunneling between the system and reservoir $\lambda$. 
    The reservoirs are taken as a set of 1D leads each connecting to the left and right boundary sites. Further we consider them to be  in the wide-band limit, such that the corresponding self-energies are frequency independent. 
    The explicit forms of these system, reservoir and tunneling Hamiltonians are provided in Appendix~\ref{sec:Floquet_NEGF}. 
    
	\subsubsection{Two-terminal conductance}
	For the transport setup described above, the steady-state current can be expressed in terms of Floquet Green's functions using the equation of motion approach for the periodically driven system and reservoir operators (see Appendix~\ref{sec:Floquet_NEGF} for details). Following standard NEGF methods, the current (recalling that we are using units with $e=1, \hbar=1$), flowing between the two reservoirs can be written as a Floquet generalization of the Landauer formula~\cite{Kohler2005},
	\begin{equation}
		I_x
		=
		\int_{-\infty}^{\infty} \frac{d\omega}{2\pi}
		\sum_{q\in\mathbb{Z}}
		\Big[
		T^{(q)}_{\rm RL}(\omega)\, f_L(\omega)
		-
		T^{(q)}_{\rm LR}(\omega)\, f_R(\omega)
		\Big],
	\end{equation}
	where $T^{(q)}_{\rm LR}(\omega)$ is the transmission probability for an electron with energy $\omega$ to be transmitted from the left to the right reservoir while absorbing ($q>0$) or emitting ($q<0$), $|q|$ photons from the periodic drive. The Floquet transmission coefficients are given by
	\begin{equation}
		\label{eq:G-transmission-tro-terminal}
		T^{(q)}_{\lambda\lambda'}(\omega)
		=
		4\pi^2
		\mathrm{Tr}\!\left[
		G^{(q)\dagger}(\omega)
		\Gamma_\lambda
		G^{(q)}(\omega)
		\Gamma_{\lambda'}
		\right],
	\end{equation}
    where we recall that $\lambda ,\lambda'\in \{L,R\}$ and  $\lambda\neq \lambda'$. 
    Eq.~\eqref{eq:G-transmission-tro-terminal} represents the Floquet extension of the standard Landauer-NEGF relation between the transmission probability and the Green's function. Here we work with one-dimensional reservoirs in the wide-band limit, for which  the effective coupling matrices are given by:  
    $\Gamma_{L,xy,x'y'}=\gamma \delta_{xx'}\delta_{yy'}\delta_{x,1}$ and $\Gamma_{R,xy,x'y'}=\gamma \delta_{xx'}\delta_{yy'}\delta_{x,2 n_x}$, with $\gamma$ as the single system-bath coupling parameter. Further details are provided in Appendix~\ref{sec:Floquet_NEGF}.

	We now apply a small bias by setting the chemical potentials of the left and right reservoirs to $\mu_L = \mu_F + \Delta\mu$ and $\mu_R = \mu_F$, where $\mu_F$ denotes the equilibrium chemical potential of the reservoirs. In the linear-response limit $\Delta\mu\to 0$, the two-terminal conductance (in units of $e^2/h$), is defined as $G_2(\mu_F)=\left.\frac{2 \pi I_x}{\Delta\mu}\right|_{\Delta\mu\to 0}$, which evaluates to
	\begin{equation}
		G_2(\mu_F)
		=
		\sum_{q\in\mathbb{Z}} T^{(q)}_{\rm LR}(\mu_F).
		\label{eq:G2}
	\end{equation}
	%~~~~~~~~~~~~~~~~~~~~~~~~~~~~~~~~~~~~~~~~~~~~~~~~~~FIG-3~~~~~~~~~~~~~~~~~~~~~~~~~~~~~~~~~~~~~~~~~~~~~~~~~~~
	\begin{figure*}[ht!]
		\centering
		\includegraphics[width=.8\linewidth]{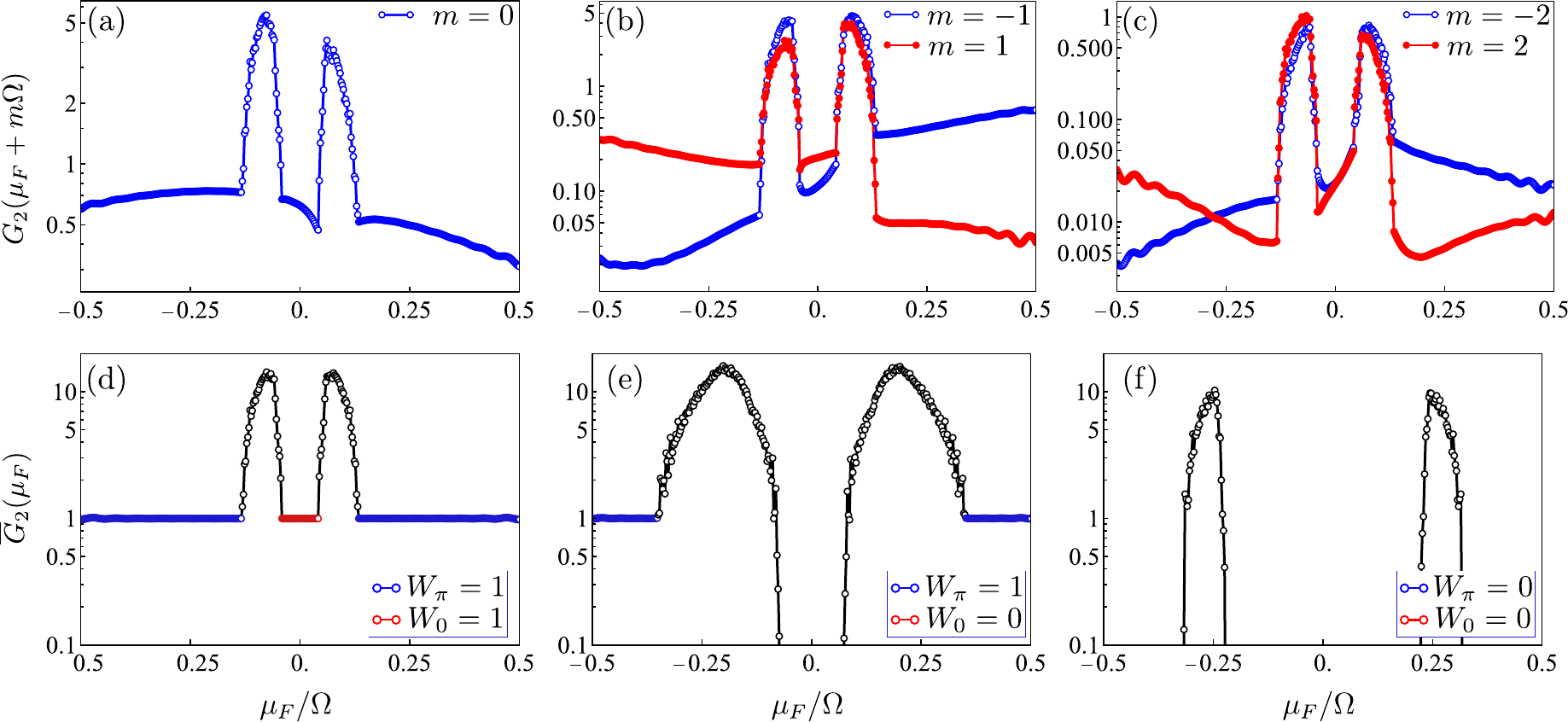}
		\caption{
        %\textbf{Photon-assisted Floquet conductance.}
        (a--c) Individual Floquet sideband contributions $G_{2}(\mu_{F}+m\Omega)$ for $m = 0, \pm1, \pm2$, shown as a function of the unbiased chemical potential $\mu_F$, computed using Eq.~\eqref{eq:G2}. (d--f) The summed dc conductance $\overline{G}_{2}(\mu_{F})$, computed using Eq.~\eqref{eq:G2_bar}, obtained by summing over $7$ sidebands ($m=-3$ to $3$), exhibits quantized plateaus determined by the Rudner winding invariants $W_{0}$ and $W_{\pi}$. 
        A horizontal solid line at $\overline{G}_2(\mu_F)=e^2/h$ is shown in panels (d--f) to highlight the quantization. 
        The system size of the underlying lattice is $n_x = 20$ and $n_y = 40$, with $\rho_{L(R)} = 1/4\pi$ and  reservoir temperature $\mathcal{T}_{L(R)} = 0$.
        All other parameters are the same as in Fig.~\ref{fig:FIG2_QES}(a) for panels (a--d), while panels (e) and (f) correspond to the parameter regimes shown in Fig.~\ref{fig:FIG2_QES}(b) and Fig.~\ref{fig:FIG2_QES}(c), respectively.}
		\label{fig:FIG4-muF}
	\end{figure*}
	A key feature of Floquet transport through topological systems is the existence of the ``Floquet sum rules'' ~\cite{Esin2018,Dehghani2015,Farrell2016,Yap2017,Bajpai2020,RudnerLindnerReview2019,Mondal2025},
    which states that the summed conductance defined as,
	\begin{equation}
		\overline{G}_2(\mu_F)
		=
		\sum_{m\in \mathbb{Z}} G_2(\mu_F + m\Omega),
		\label{eq:G2_bar}
	\end{equation}
    shows quantization. The summed conductance therefore provides a robust two-terminal transport signature of Floquet topological systems.
	
    \subsubsection{Longitudinal and transverse bond currents}
    In the strip geometry (recall Fig.~\ref{fig:FIG-SETUP}), in addition to the two-terminal conductance, we now  compute the local bond currents along the longitudinal ($x$) and transverse ($y$) directions of the lattice. This will help to better characterize the spatial structure of transport. For a time-periodic Hamiltonian satisfying $H(t+T)=H(t)$, the local particle density operator at each site $\mathbf r$ of the lattice is defined as $\mathcal{N}(\mathbf r,t)=c^\dagger(\mathbf r,t)c(\mathbf r,t)$, here $c(\mathbf r,t)$ denotes the basis introduced in Sec.~\ref{sec:Floquet topological phases}. However unlike in Sec.~\ref{sec:Floquet topological phases}, we use a different labeling  $\mathbf r=(x,y)$ with $x=n_L+1,\ldots,n_L+2n_x$ and $y=1,\ldots,n_y$. Using the Heisenberg equation of motion together with the continuity equation, one can define the instantaneous bond current $J_{\mathbf r\to\mathbf r'}(t)$ flowing from site $\mathbf r$ to a neighboring site $\mathbf r'$. The time-averaged bond current  is  given by
	\begin{eqnarray}
	    \mathcal{J}_{\mathbf r\to\mathbf r'}
		&=&
		\frac{1}{T}\int_0^T dt\, J_{\mathbf r\to\mathbf r'}(t)
		\nonumber\\
		&=&
		\frac{2}{T}\int_0^T dt\,
		\mathrm{Im}
		\!\left[
		H_{\mathbf r,\mathbf r'}(t)
		\left\langle
		c^\dagger(\mathbf r,t)c(\mathbf r',t)
		\right\rangle
		\right],\label{eq:J_rr'}
	\end{eqnarray}
    where $H_{\mathbf r,\mathbf r'}(t)$ denotes the hopping matrix between lattice sites $\mathbf r$ and $\mathbf r'$ appearing in the system Hamiltonian in Eq.~\eqref{eq:h5step_strip} and {the expectation value $\langle\cdots\rangle$  is obtained using the Floquet--NEGF framework. We again consider zero temperature case with the lead chemical potentials set at $\mu_L=\mu_F+\Delta \mu$ and $\mu_R=\mu_F$. Let us further define the net  currents, $\mathcal{J}_x$ and $\mathcal{J}_y$,  along the $x$- and $y$-directions, respectively by summing  local bond currents over all rows or all  columns (within the system).  Thus 
\begin{align}
\mathcal{J}_x(\mu_F)&=\sum_{y=1}^{n_y} \mathcal{J}_{x,y\to x+1,y} (\mu_F)\\
\mathcal{J}_y(\mu_F)&=\sum_{x=n_L+1}^{n_L+2n_x} \mathcal{J}_{x,y\to x,y+1}(\mu_F)
\end{align}
    }
 The linear response time-averaged bond conductances  (in units of $e^2/h$), are then given by
    \begin{align}
    \mathcal{G}_{x(y)}(\mu_F)
	=
	\lim_{\Delta\mu\to0}
	\frac{2\pi\mathcal{J}_{x(y)}(\mu_F)}{\Delta\mu}.
    \end{align}
    After some computations we get   the explicit forms,
	\begin{subequations}
		\begin{align}
			\mathcal{G}_x(\mu_F)
			&=
			\sum_{y=1}^{n_y}
			\sum_{q,\alpha,\beta}
			4\pi\,\mathrm{Im}
			\Big[
			H^{(q)}_{(x,y),(x+1,y);\beta\alpha}
			\,
			\mathcal{F}^{(q)}_{\alpha\beta}(\mu_F)
			\Big],
			\label{eq:bond-current-xy-a}
			\\
			\mathcal{G}_y(\mu_F)
			&=
			\sum_{x=n_L+1}^{n_L+2 n_x}
	\sum_{q,\alpha,\beta}
			4\pi\,\mathrm{Im}
			\Big[
			H^{(q)}_{(x,y),(x,y+1);\beta\alpha}
			\,
			\mathcal{F}^{(q)}_{\alpha\beta}(\mu_F)
			\Big],
			\label{eq:bond-current-xy-b}
		\end{align}
	\end{subequations}
	where $\alpha,\beta$ label Floquet quasienergy eigenstates, $(x,y)$ denotes lattice sites, $H^{(q)}$ represents the $q$-th Floquet harmonic of the Hamiltonian, and $\mathcal{F}^{(q)}_{\alpha\beta}(\mu_F)$ (defined in Appendix~\ref{sec:bond_currents_using_negf}) denotes the nonequilibrium Floquet cross-occupation matrix. A detailed derivation of this conductance expressions are  provided in Appendix~\ref{sec:bond_currents_using_negf}. 
	
	In the strip geometry, current conservation implies  that the longitudinal current must be identical across any cross-section of the strip,
    hence $\mathcal{G}_{x}(\mu_F)$ must be independent of $x$ and equal to $G_2(\mu_F)$.  The two-terminal summed conductance defined as 
	\begin{equation}
		\overline{\mathcal{G}}_{2,x}(\mu_F)
		=
		\sum_{m\in\mathbb{Z}}
		\mathcal{G}_x(\mu_F+m\Omega).
		\label{eq:X netJbulk1}
	\end{equation}
	is thus expected to be identical to the 
    summed conductance defined in Eq.~\eqref{eq:G2_bar}.
    %~~~~~~~~~~~~~~~~~~~~~~~~~~~~~~~~~~~~~~~~~~~~~~~~~~FIG-4~~~~~~~~~~~~~~~~~~~~~~~~~~~~~~~~~~~~~~~~~~~~~~~~~~~
	\begin{figure*}[ht!]
		\centering
		\includegraphics[width=0.9\linewidth]{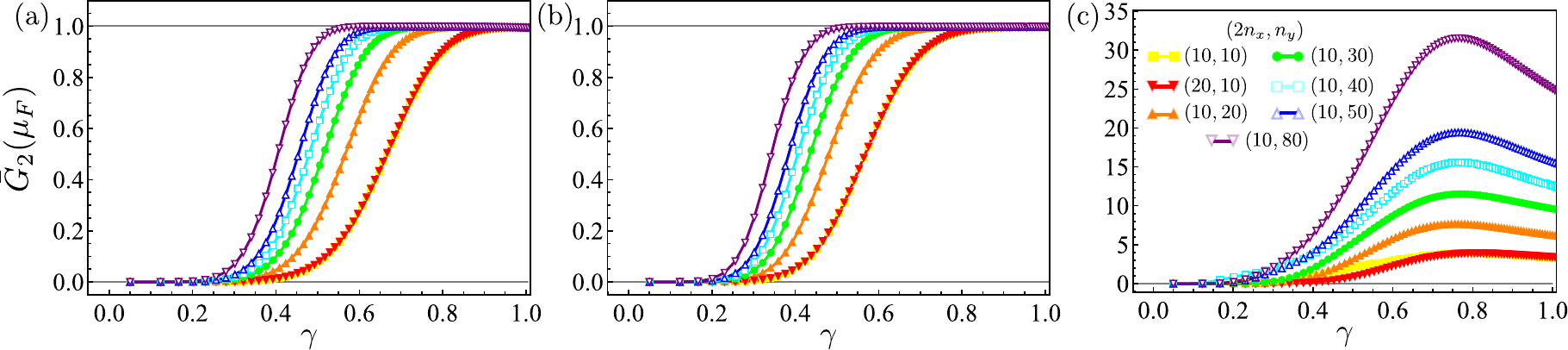}
		\caption{
			Zero-bias conductance as a function of the system--reservoir coupling $\gamma$. 
			The reservoir chemical potential is set to (a) $\mu_F=\Omega/2$, (b) $\mu_F=0$, and (c) $\mu_F=0.075\Omega$.
			All other parameters are the same as in Fig.~\ref{fig:FIG2_QES}(c), which corresponds to the anomalous Floquet topological phase. 
		}
		\label{fig:FIG5-COUPLING}
	\end{figure*}
	% -------------------------------------------------------------
	
    The transverse bond conductance $\mathcal{G}_{y}(\mu_F)$ will generally exhibit spatial variations ($y$-dependence) near the sample edges but approaches a constant value sufficiently deep in the bulk. The net Hall conductance of the system can therefore be extracted from the bulk value of the transverse conductance, leading to \begin{equation}
		\overline{\mathcal{G}}_{H,y}(\mu_F)
		=
		\sum_m
		\mathcal{G}_y(\mu_F+m\Omega),
		\label{eq:Y netJbulk2}
	\end{equation}
    where $y$ can be chosen as any bulk point (i.e, far from the edges).  
     The above expressions for local conductances continue to hold in the cylindrical geometry
where, however,  we can show, translational invariance along the $y$ direction renders uniform $\overline{\mathcal{G}}_{y}(\mu_F)$  across the system, and this allows us to obtain a  simplified  expression of the net Hall conductance $\overline{\mathcal{G}}_{H,y}(\mu_F)$ (see Sec.~\eqref{sec: Hall conductance quantization in a cylindrical geometry}). 

In the following section we present numerical results for the two-terminal (longitudinal) conductance and transverse bond conductance obtained within this formalism.
    
	\section{Numerical results} \label{sec:Numerical results}
	\subsection{Quantization in a strip geometry} \label{sec:strip geometry}
	In this section we present numerical results for two-terminal transport in the setup shown in Fig.~\ref{fig:FIG-SETUP}. We focus on the three parameter regimes considered in  Fig.~\ref{fig:FIG2_QES}(a)--(c) for which the periodically driven central region hosts topological edge states either at both $0$ and  $\Omega/2$ or has edge states at quasienergy $\Omega/2$ or 
    has  no topological edge states.
    
    We first consider the anomalous Floquet phase shown in Fig.~\ref{fig:FIG2_QES}(a), which supports chiral edge modes at both $0$ and $\Omega/2$ quasienergies. This phase is characterized by winding numbers $(W_0,W_{\pi})=(1,1)$, while all Floquet bulk band Chern numbers vanish, reflecting its anomalous Floquet topological phase.
	For this anomalous phase, the top row of Fig.~\ref{fig:FIG4-muF} shows the zero-bias conductance $G_{2}(\mu_F+m\Omega)$, computed using Eq.~\eqref{eq:G2}. The  Figures~\ref{fig:FIG4-muF}(a--c) show the contribution to the conductance from different Floquet zones: panel (a) corresponds to the central zone ($m=0$), panel (b) to the first upper ($m=1$) and lower ($m=-1$) zones, and panel (c) to the second upper ($m=2$) and lower ($m=-2$) zones. As expected in Floquet systems, the conductance within individual Floquet zones is generally not quantized. However, the summed conductance $\overline{G}_{2}(\mu_F)$ as shown in Fig.~\ref{fig:FIG4-muF}(d), exhibits robust quantization. In fact we find that the convergence of the sum is quite fast and in Fig.~\ref{fig:FIG4-muF}(d), the number of terms included was $7$ ($m=-3$ to $3$).
        In this phase the quantized plateau heights correspond to the absolute values of the Floquet winding invariants and satisfy
	\begin{equation}
		\overline{G}_{2}(0)=|W_0|,
		\qquad
		\overline{G}_{2}(\Omega/2)=|W_{\pi}|,
	\end{equation}
	in units of $e^2/h$, reflecting the contributions of the chiral edge modes at quasienergies $0$ and $\Omega/2$.

    Similar to the Floquet anomalous topological phase, for parameters corresponding to the phase shown in Fig.~\ref{fig:FIG2_QES}(b), where $W_{\pi}=1$, a quantized conductance plateau appears when $\mu_F$ lies in the quasienergy window of the $\Omega/2$ edge mode, as shown in Fig.~\ref{fig:FIG4-muF}(e).
	Finally, we also examine the trivial gapped phase that does not host any Floquet edge states, as shown in Figs.~\ref{fig:FIG2_QES}(c).  In this trivial regime, characterized by $(W_0,W_{\pi})=(0,0)$, the conductance vanishes whenever $\mu_F$ lies within the bulk gap, as shown in Fig.~\ref{fig:FIG4-muF}(f). 
	
    Next we discuss how the conductance quantization shown in Fig.~\ref{fig:FIG4-muF}  depends on the system-reservoir couplings and the system sizes along the $x$- and $y$- directions.  The dependence of the conductance on the system–reservoir coupling parameter $\gamma$, for different values of $n_x$ and $n_y$, is shown in Fig.~\ref{fig:FIG5-COUPLING}, for  parameters corresponding to the Floquet anomalous phase in  Fig.~\ref{fig:FIG2_QES}(a). Panels~(a--c) plot the summed conductance as a function of $\gamma$ for three  chemical potentials: $\mu_F=\Omega/2$, $\mu_F=0$, and $\mu_F=0.07\,\Omega$, respectively. The first two values probe transport dominated by the $\Omega/ 2$ and $0-$ Floquet edge modes, while the third lies within the bulk continuum and captures contributions from bulk states.	%~~~~~~~~~~~~~~~~~~~~~~~~~~~~~~~~~~~~~~~~~~~~~~~~~~FIG-4~~~~~~~~~~~~~~~~~~~~~~~~~~~~~~~~~~~~~~~~~~~~~~~~~~~
	\begin{figure*}[ht!]
		\centering
		\includegraphics[width=0.8\linewidth]{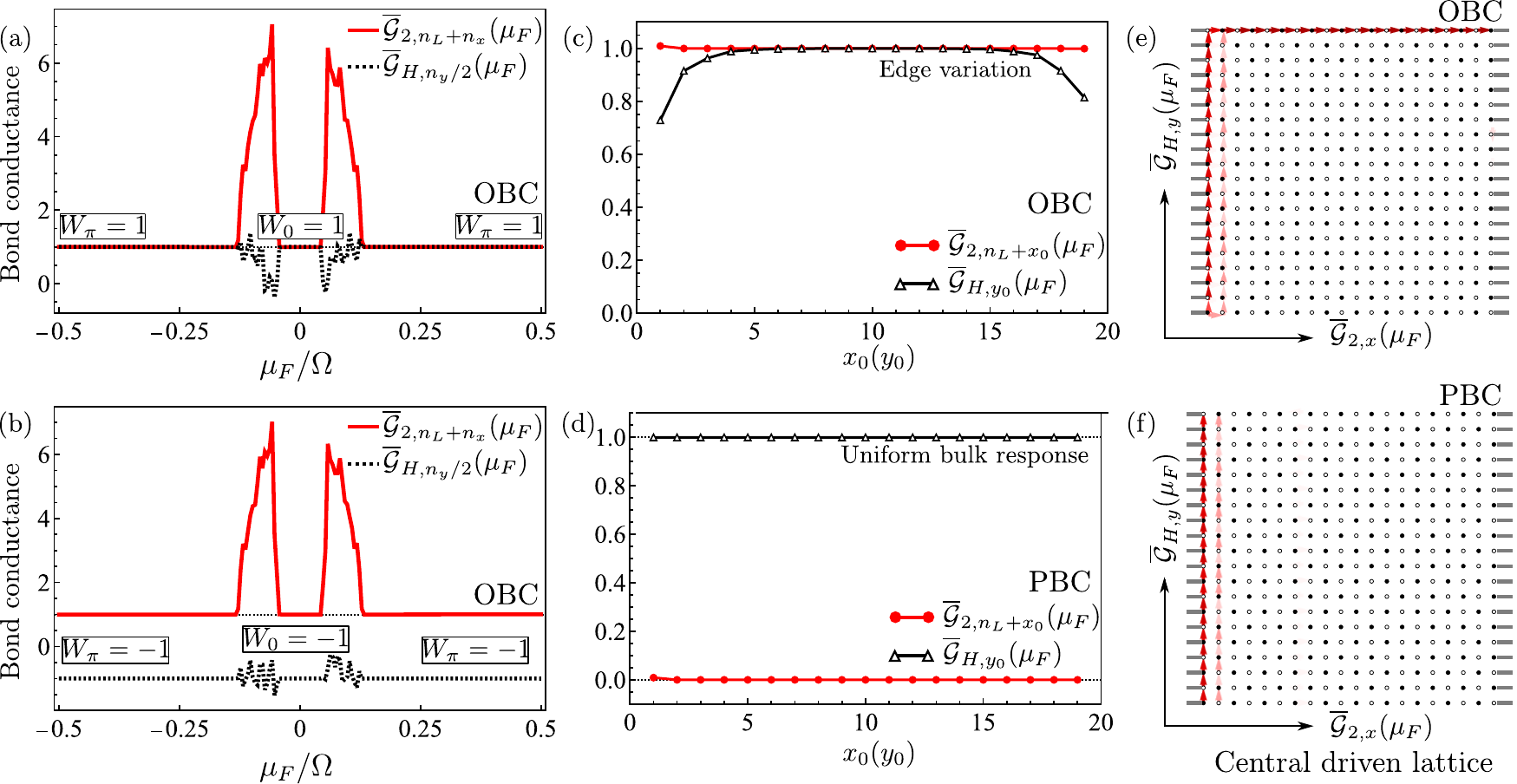}
		\caption{Longitudinal ($\overline{\mathcal{G}}_{2,n_L+n_x}(\mu_F)$, solid lines) and transverse ($\overline{\mathcal{G}}_{H,n_y/2}(\mu_F)$, dashed lines) conductances for the anomalous Floquet phase with winding numbers (a) $(W_0,W_{\pi})=(1,1)$ and (b) $(W_0,W_{\pi})=(-1,-1)$. (c,d) Spatial profiles of the bond conductances $\overline{\mathcal{G}}_{2,n_L+x_0}(\mu_F)$ and $\overline {\mathcal{G}}_{H,y_0}(\mu_F)$ as a function of $x_0$ and $y_0$ in the system, respectively, 
        for a $(2n_x,n_y)=(20,20)$ system under open boundary conditions (OBC) and periodic boundary conditions (PBC), respectively. (e,f) Distributions of the bond conductances $\overline{\mathcal{G}}_{2,x} (\mu_F)$ versus $\overline{\mathcal{G}}_{H,y}(\mu_F)$ in the central driven lattice under OBC and PBC, respectively.}
		\label{fig:FIG_G2GH_bulk}
	\end{figure*}
	%~~~~~~~~~~~~~~~~~~~~~~~~~~~~~~~~~~~~~~~~~~~~~~~~~~FIG-4~~~~~~~~~~~~~~~~~~~~~~~~~~~~~~~~~~~~~~~~~~~~~~~~~~~
	We find that for relatively small system sizes (e.g., $n_x = n_y \lesssim 20$), conductance quantization is observed only when the coupling to the reservoirs is sufficiently strong. As the transverse system size ($n_y$) is increased, quantization emerges even in the weak-coupling regime. This trend is clearly visible in Figs.~\ref{fig:FIG5-COUPLING}(a,b), where $\mu_F$
    is $\Omega/2$ and $0$, respectively. We also notice the same trend as long as $\mu_F$ is within the window of $0$ and $\Omega/2$ edge states. 
    We also find that
    increasing the longitudinal system size ($n_x$) has a weaker effect and one quickly attains saturation, 
    i.e., conductance reaches to the quantizated value already at rather small value of $n_x$. This is shown in Figs.~\ref{fig:FIG5-COUPLING}(a,b), where we observe that for $n_x=5$ and $n_x=10$ conductance has similar dependence on the coupling parameter, for a fixed value of $n_y=10$. Thus we see that we can obtain quantization for arbitraritly small value of $\gamma$ as long as we correspondingly take $n_y$ to be sufficiently large. 
     In this limit, the larger number of boundary sites increases the total coupling between the reservoirs and the chiral edge channel, thereby facilitating efficient injection into the edge mode. As a result, conductance quantization is restored even for relatively weak system--reservoir coupling. 
	
	By contrast, when $\mu_F$ lies within the bulk bands, as shown in Fig.~\ref{fig:FIG5-COUPLING}(c), the conductance does not saturate to a quantized value but instead increases monotonically with the coupling strength. In this regime, increasing the transverse system size simply introduces additional bulk transport channels, resulting in a larger but nonquantized conductance. 

	To gain further insight into the spatial structure of transport, we next examine the net current flow within the driven central region. Using the Floquet nonequilibrium Green's-function (NEGF) formalism, we compute the longitudinal and transverse conductances from bond currents averaged over one driving period along the $x$ and $y$ directions, respectively, as defined in Eqs.~(\ref{eq:X netJbulk1})--(\ref{eq:Y netJbulk2}).
    Figure~\ref{fig:FIG_G2GH_bulk} summarizes the bulk and spatially resolved conductance properties obtained within the Floquet--NEGF formalism. Panel~(a) shows the bulk longitudinal and transverse conductances as functions of the unbiased chemical potential $\mu_F$ for the Floquet topological phase characterized by $(W_{0},W_{\pi})=(1,1)$. We observe that both the longitudinal conductance $\overline{\mathcal{G}}_{2,n_L+n_x}(\mu_F)\approx|W_{\epsilon}|$ and the transverse (Hall) conductance $\overline{\mathcal{G}}_{H,n_y/2}(\mu_F)\approx W_{\epsilon}$ exhibit quantized plateaus when $\mu_F$ lies inside the quasienergy gaps of the driven system, as shown in Fig.~\ref{fig:FIG_G2GH_bulk}(a). The plateau values are determined by the Floquet winding invariants $W_{\epsilon}$ for $\epsilon=0$ and $\epsilon=\pi$, consistent with the winding numbers independently computed from the Floquet time-evolution operator (for details, see Appendix~\ref{sec: Floquet winding invariants}).

    Importantly, unlike the longitudinal conductance, the transverse conductance $\overline{\mathcal{G}}_{H,y_0}(\mu_F)$ reflects both the magnitude and the sign of the winding invariants and therefore encodes the chirality of the Floquet edge modes. To demonstrate this, we reverse the driving sequence in the Floquet protocol; i.e., in Eq.~\eqref{eq:h5step_strip}, $j$ varies from $5$ to $1$. This reversal flips the signs of the Floquet winding invariants, yielding $(W_0,W_{\pi})=(-1,-1)$. The corresponding bond-conductance results are shown in Fig.~\ref{fig:FIG_G2GH_bulk}(b). As expected, reversing the driving sequence changes the sign of $\overline{\mathcal{G}}_{H,n_y}(\mu_F)$, while the longitudinal conductance $\overline{\mathcal{G}}_{2,n_L+n_x}(\mu_F)$ remains unchanged, reflecting its dependence on $|W_\varepsilon|$ rather than $W_\varepsilon$, with the latter encoding the chirality of the edge modes.
    
    To understand the spatial origin of these quantized responses, we next examine the spatially resolved bond conductances within the driven lattice. Fig.~\ref{fig:FIG_G2GH_bulk}(c) shows the spatial profiles of the longitudinal and transverse bond conductances for a system of size $(2n_x,n_y)=(20,20)$. We find that $\overline{\mathcal{G}}_{2,n_L+x_0}(\mu_F)$ as a function of $x_0$ which varies from $1$ to $2n_x$ and we observe that it, remains essentially uniform throughout the system, reflecting conservation of longitudinal current in the steady state. In contrast, we observe $\overline{\mathcal{G}}_{H,y_0}(\mu_F)$ with respect to $y_0$, which varies from $1$ to $n_y$, exhibits pronounced spatial variations near the boundaries of the strip, but rapidly approaches a quantized value in the bulk. Implementing periodic boundary conditions (PBC), as shown in Fig.~\ref{fig:FIG_G2GH_bulk}(d), restores translational invariance along the transverse direction. As a result, the longitudinal bond conductance $\overline{\mathcal{G}}_{2,n_L+x_0}(\mu_F)$ vanishes identically, while the transverse bond conductance $\overline{\mathcal{G}}_{H,y_0}(\mu_F)$ becomes spatially uniform across the system and correctly reproduces both the magnitude and the sign of the Floquet winding invariants. In this geometry, the Hall response arises purely from the bulk and attains its quantized value at every site.
	
    The distinction between open and periodic boundary conditions is further illustrated by the bond-conductance distributions in the central lattice i.e. $x=n_L+1,\ldots, n_L+2n_x$ and $y=1,\ldots, n_y$, shown in Figs.~\ref{fig:FIG_G2GH_bulk}(e) and~\ref{fig:FIG_G2GH_bulk}(f), respectively. When $\mu_F$ lies within the range of topological edge modes under OBC, both longitudinal and transverse conductance flow predominantly along the system edges, as shown in Figs.~\ref{fig:FIG_G2GH_bulk}(e), highlighting the effectively one-dimensional nature of edge-state transport. Under PBC, by contrast, the longitudinal conductance vanishes, while the Hall conductance remains quantized and spatially uniform along the $y$-direction, yet localized near the left boundary, as seen in Fig.~\ref{fig:FIG_G2GH_bulk}(f).
    
	At this stage, it is useful to clarify the conceptual distinction between the strip and cylindrical geometries. In the strip setup, conductance quantization arises from chiral edge modes that couple directly to the reservoirs and dominate transport. The quantized longitudinal and Hall responses therefore reflect edge-state–mediated current flow. In contrast, in the cylindrical geometry with periodic boundary conditions along $y$, translational invariance is restored and the Hall conductance becomes spatially uniform along $y$, reflecting a purely bulk topological response. The quantized value is then determined directly by the Floquet winding invariants. These two perspectives are fully consistent: the edge-dominated transport in the strip geometry and the uniform bulk response in the cylinder represent complementary manifestations of Floquet bulk--boundary correspondence.
    %~~~~~~~~~~~~~~~~~~~~~~~~~~~~~~~~~~~~~~~~~~~~~~~~~~FIG-4~~~~~~~~~~~~~~~~~~~~~~~~~~~~~~~~~~~~~~~~~~~~~~~~~~~
        	\begin{figure}[ht!]
        		\centering
        		\includegraphics[width=0.8\linewidth]{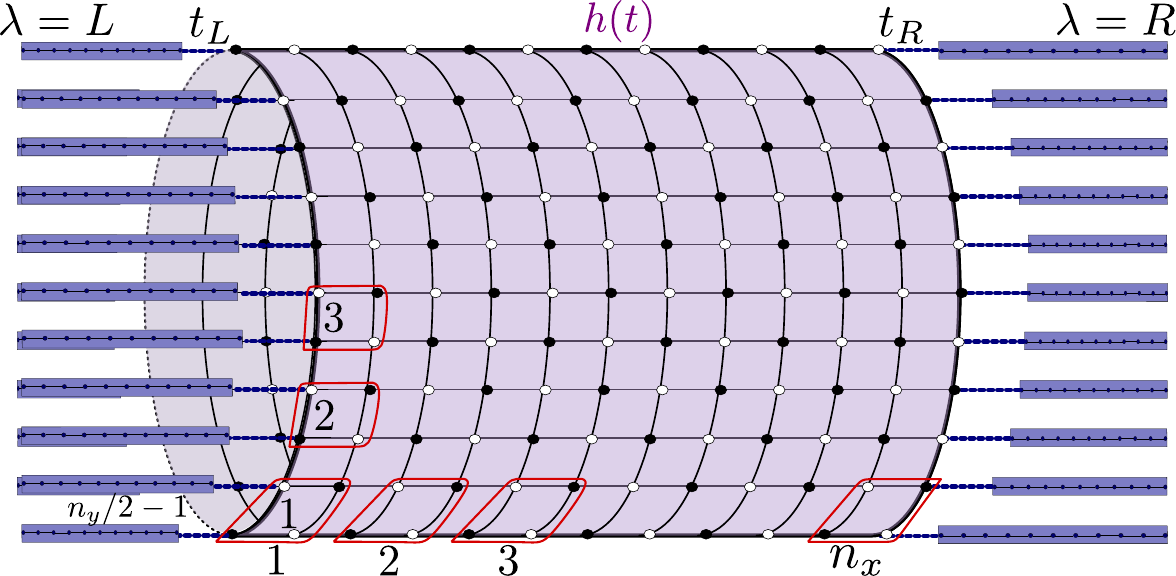}
        		\caption{System setup: The system described by Eq.~\eqref{eq:H5_cylinder}. The ends of the cylinder are connected to two non-driven wide-band Fermionic reservoirs. The cylinder is finite along the $x$-direction, with a total of $2 n_x$ sites, and $k_y$ is a good quantum number. The lengths of the left and right reservoirs are denoted by $n_L$ and $n_R$, respectively. The symbolic red square boxes denotes the cells in both $x$ and $y$ direction.}
        		\label{fig:cylindrical_geo}
        	\end{figure}
        	%~~~~~~~~~~~~~~~~~~~~~~~~~~~~~~~~~~~~~~~~~~~~~~~~~~FIG-4~~~~~~~~~~~~~~~~~~~~~~~~~~~~~~~~~~~~~~~~~~~~~~~~~~~
	\begin{figure*}[ht!]
		\centering
		\includegraphics[width=0.8\linewidth]{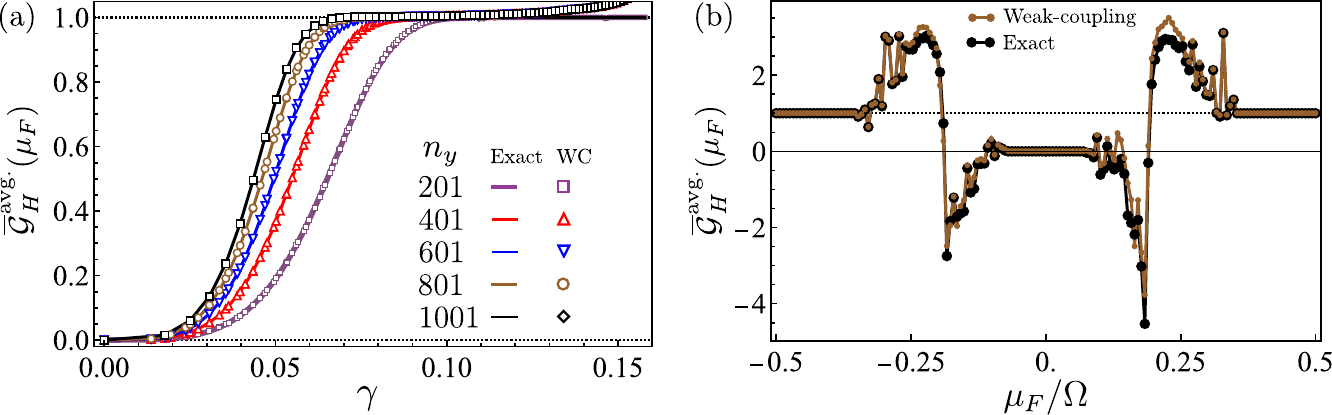}
        \caption{(a) Comparison of the exact Floquet--NEGF Hall conductance (solid lines) with the weak-coupling (WC) approximation (markers) as a function of the system--lead coupling $t_L$, for different momentum resolutions by varying $n_y$. Each $n_y$ is represented by a distinct color shared by the corresponding exact and weak-coupling curves. (b) Hall conductance as a function of the unbiased chemical potential, for a fixed value of coupling $t_L=1$, computed using the exact and weak-coupling approaches. Here the parameters are kept same as Fig.~\ref{fig:FIG2_QES}(b).
         \label{fig:GH_weak_coupling_GvsmuF}
        }
    \end{figure*}
    
	Motivated by the strong transverse-size dependence of quantization in the strip geometry shown in Fig.~\ref{fig:FIG5-COUPLING}, we  analyze, in the next section, the Hall conductance for the same model in a cylindrical setup. The corresponding Hamiltonian is given in Eq.~\eqref{eq:H5_cylinder}, and the setup is illustrated in Fig.~\ref{fig:cylindrical_geo}. In this configuration, periodic boundary conditions along the $y$-direction render $k_y = \frac{2\pi p}{n_y}$, where $p = 0, \ldots, n_y - 1$, which significantly simplifies the analytical expressions. In this setup, the current expressions reduce to a simpler form, enabling a transparent analytical connection between bulk Floquet topology and quantized transport (see Appendix~\ref{sec:bond_currents_using_negf} for details).
    
    \subsection{Quantization of Hall conductance in a cylindrical geometry}\label{sec: Hall conductance quantization in a cylindrical geometry}
    In this section, we analyze the Hall conductance in a cylindrical geometry. In the weak system-reservoir coupling regime, the Hall response is fully determined by the steady-state density matrix of the periodically driven system~\cite{Seradjeh_PRL,Kumari2024,FFL2024}, establishing a direct connection to Floquet conductance sum rules.

    For the cylindrical geometry, we can repeat the computations of local bond currents and obtain the expression   given in Eq.~\eqref{eq:bond-current-xy-b}. 
    From symmetry it is clear that the $y$-current is independent of the spatial $y$-position. 
   We can then define the spatially and time-averaged summed Hall conductance as,
    \begin{equation}
        \overline{\mathcal{G}}^{\mathrm{avg}}_{H}(\mu_F)=\frac{1}{n_y}\sum_{y=1}^{n_y}\sum_{m}\mathcal{G}_{H,y}(\mu_F+m\Omega).
    \end{equation}
   As discussed in Sec~\ref{sec:Floquet topological phases}, using the translation invariance in the $y$ direction, the system is effectively described by disconnected 1D chains and we denote the $n$-th Fourier mode of the Hamiltonian  by $H^{(n)}(k_y)$, with components $H^{(n)}_{x,x'}(k_y)$ ($x,x'=1,\ldots,2n_x$).  
    Using Eq.~\eqref{eq:bond-current-xy-b}, and translation invariance in the $y$-direction,  this can be expressed in momentum space:
    \begin{equation}
\overline{\mathcal{G}}^{\text{avg.}}_H(\mu_F)
    	=
    	\frac{2\pi}{n_y}
    	\sum_{k_y}
    	\sum_{q,\alpha,\beta}
    	2\,\mathrm{Im}
    	\!\left[
    	\mathcal{J}^{(q)}_{k_y,\beta\alpha}\,
    	\bar{\mathcal{F}}^{(q)}_{k_y,\alpha\beta}(\mu_F)
    	\right],
    	\label{eq:Ghall-no-weak}
    \end{equation}
    where $k_y$ denotes the conserved transverse momentum, and $\alpha,\beta$ label the Floquet quasienergy bands and the quantities $\mathcal{J}^{(q)}_{k_y,\beta\alpha}$ and $\bar{\mathcal{F}}^{(q)}_{\alpha\beta}$ are defined as
    \begin{subequations}
    	\begin{align}
    		\mathcal{J}^{(q)}_{k_y,\beta\alpha}
    		&=
    		\sum_{l,k}
    		\left\langle
    		u_{\beta}^{-\,(l)}
    		\left|
    		\frac{\partial H^{(l-k+q)}}{\partial k_y}
    		\right|
    		u_{\alpha}^{-\,(k)}
    		\right\rangle,
    		\label{eq:no-weak-J}
    		\\
    		\bar{\mathcal{F}}^{(q)}_{k_y,\alpha\beta}(\mu_F)
    		&=
    		\sum_{m,n}\left\langle
    			u_{\alpha}^{+(n+q)}
    			\left|
    			\Gamma_L
    			\right|
    			u_{\beta}^{+(n)}
    			\right\rangle\\&\times
    		\frac{
    			1
    		}{
    			\big(\mu_F^{(m)}-e_{\alpha}^{(n)}+i\gamma_\alpha\big)
    			\big(\mu_F^{(m)}-e_{\beta}^{(n+q)}-i\gamma_\beta\big)
    		}\nonumber.
    		\label{eq:no-weak-F}
    	\end{align}
    \end{subequations}
    Here $\mu_F^{(n)}=\mu_F+n\Omega$, 
    $|u_{\alpha}^{+(n)}\rangle$ and  $|u_{\alpha}^{-(n)}\rangle$ are the Floquet eigenmodes of the non-hermitian system $H(t)+i\Gamma$ and $H(t)-i\Gamma$, respectively, in the extended space and $e_{\alpha}^{(n)}+i\gamma_\alpha$ and $e_{\alpha}^{(n)}-i\gamma_\alpha$ are the  corresponding quasienergies. Also, since we now have 1D chains, we take $\Gamma=\Gamma_L+\Gamma_R$, where $\Gamma_{L,x,x'}=\gamma\delta_{x,x'}\delta_{x,1}$ and $\Gamma_{R,x,x'}=\gamma\delta_{x,x'}\delta_{x,2n_x}$. Additionally,  $e_{\alpha}^{(n)}=e_{\alpha}+n\Omega$. For simplicity we have suppressed above the $k_y$ dependence from $H^{(l)}, |u_{\alpha}\ra, e_{\alpha}$ and $ \gamma_{\alpha}$

    As discussed for  the strip geometry in Sec.~\eqref{sec:strip geometry}, the Hall conductance will show quantization in the weak-coupling regime as long as considers sufficiently  large system $n_x$ and $n_y$. We use this observation to obtain an analytical proof in the combined limits of weak system--reservoir coupling and large  system size.
    
   In the weak-coupling limit, the spectra of the Hermitian (isolated) and non-Hermitian (reservoir-coupled) Floquet systems are related perturbatively~\cite{Bukov2015,Rodriguez-Vega_2018}. 
   Let us denote by $\epsilon_{\alpha}$, and $|\phi_{\alpha}(t)\rangle$,  the quasi-energy and Floquet modes of the isolated  Hamiltonian in the cylindrical geometry which are thus similar to those defined, for the torus, in   Eqs.~(\ref{eq:eigen_value_eq_Floquet_mods},\ref{eq:floqfour}). The isolated system wavefunctions are now defined in the site basis. 
   The reservoir coupling enters via the self-energy $i\Gamma$, with $\Gamma=\Gamma_L+\Gamma_R$, acting as a small non-Hermitian perturbation. The corresponding bi-orthogonal Floquet eigenstates 
   \begin{equation}
      \{|u^{-}_\alpha(t)\rangle,|u^{+}_\alpha(t)\rangle\} \Rightarrow |\phi_{\alpha}(t)\rangle + \mathcal{O}(\Gamma)
      \label{eq:u_weak_coupl_limit-1},
   \end{equation}
   with the real part of quasienergies is $e_{\alpha}\approx \epsilon_{\alpha}$, and the imaginary part is given by
    \begin{align}
    	\gamma_{\alpha} \approx \int_0^{T}\,\frac{dt}{T}\,\langle \phi_{\alpha}(t) |\pi \Gamma | \phi_{\alpha}(t) \rangle&
        = \sum_{p\in\mathbb{Z}}\langle \phi_{\alpha}^{(p)} |\pi \Gamma | \phi_{\alpha}^{(p)} \rangle.
        \label{eq:expan}
    \end{align}

    In this limit, each Floquet level is weakly broadened and described by a narrow Lorentzian.
    Transport is then dominated by resonant ($\alpha=\beta$) Floquet states, while interband ($\alpha\neq\beta$) and higher-harmonic ($q\neq0$) processes are suppressed. Accordingly, we decompose Eq.~\eqref{eq:Ghall-no-weak} as
    \begin{equation}
    	\overline{\mathcal{G}}^{\text{avg.}}_H(\mu_F)
    	= \overline{\mathcal{G}}_H^{(0)}(\mu_F) + \overline{\mathcal{G}}_H^{(\mathrm{off})}(\mu_F),
    \end{equation}
   	where
	\begin{equation}
		\overline{\mathcal{G}}_H^{(0)}(\mu_F)
		=\frac{2\pi}{n_y}
		\sum_{k_y,\alpha}
		2\,\mathrm{Im}
		\!\left[
		\mathcal{J}^{(0)}_{k_y,\alpha\alpha}\,
		\bar{\mathcal{F}}^{(0)}_{k_y,\alpha\alpha}(\mu_F)
		\right],
		\label{eq:Ghall-diag_q0}
	\end{equation}
	and 
	\begin{equation}
		\overline{\mathcal{G}}_H^{(\mathrm{off})}(\mu_F)
		=\frac{2\pi}{n_y}
		\sum_{k_y}
		\sum_{q,\alpha\neq\beta}
		2\,\mathrm{Im}
		\!\left[
		\mathcal{J}^{(q)}_{k_y,\beta\alpha}\,
		\bar{\mathcal{F}}^{(q)}_{k_y,\alpha\beta}(\mu_F)
		\right].
		\label{eq:Ghall-off}
	\end{equation}
	 In the weak-coupling limit it has been argued~\cite{Seradjeh_PRL,Kumari2024} that the off-diagonal and higher harmonic terms can be neglected and we need to retain only $\overline{\mathcal{G}}_H^{(0)}$. 
    In Fig.~\ref{fig:GH_weak_coupling_GvsmuF}(a) 
    we verify that this weak coupling approximation is quite accurate and reproduces the exact Hall conductance for sufficiently large $n_y$,  confirming that off-diagonal contributions are negligible. The resulting conductance plateaus are fully captured within this approximation, as shown in Fig.~\ref{fig:GH_weak_coupling_GvsmuF}(b). In this limit, we get
   \begin{align}
    	\overline{\mathcal{G}}^{\text{avg.}}_H(\mu_F)
    	\!= \overline{\mathcal{G}}_H^{(0)}(\mu_F)
        \!=\frac{4\pi}{n_y}
		\sum_{k_y,\alpha}
		\mathrm{Im}
		\,\left[
		\mathcal{J}^{(0)}_{k_y,\alpha\alpha}\,
		\bar{\mathcal{F}}^{(0)}_{k_y,\alpha\alpha}(\mu_F)
		\right]\label{eq:Ghall-diag_q0-1}.
   \end{align}
    We now focus on the diagonal contribution, i.e. Eq.~\eqref{eq:Ghall-diag_q0-1} and show that it can be simplified and leads to  analytic predictions. First, using the Floquet Hellmann--Feynman relation, the current matrix element can be written as
    \begin{align}
        \mathcal{J}_{k_y,\alpha\alpha}^{(0)} &=\sum_{l,k}
        \left\langle
        \phi_{\beta}^{(l)}
        \left|
        \frac{\partial H^{(l-k)}}{\partial k_y}
        \right|
        \phi_{\alpha}^{(k)}
        \right\rangle
        =\partial_{k_y}\epsilon_\alpha(k_y).\label{eq:Jdiag_simple}
    \end{align}
    
    Similarly, the term defined in Eq.~\eqref{eq:Ghall-diag_q0-1}, 
     \begin{equation}
     \bar{\mathcal{F}}^{(0)}_{k_y,\alpha\alpha}(\mu_F)=\sum_{m,n}
    		\frac{
    			\left\langle
    			u_{\alpha}^{+(n)}
    			\left|
    			\Gamma_L
    			\right|
    			u_{\alpha}^{+(n)}
    			\right\rangle
    		}{
            (\mu_F^{(m)}-\epsilon_{\alpha}^{(n)})^2+\gamma_\alpha^2
    		}.
     \end{equation}
     
     \noindent The above expression using the weak coupling expansion given in Eq.~\eqref{eq:u_weak_coupl_limit-1}, can be written as
    \begin{align}\nonumber
    \bar{\mathcal{F}}^{(0)}_{k_y,\alpha\alpha}(\mu_F)
        =&
        \sum_{m,n}
        \frac{
            \left\langle
            \phi_{\alpha}^{(n)}
            \left|
            \Gamma_L
            \right|
            \phi_{\alpha}^{(n)}
            \right\rangle
        }{
           (\mu_F^{(m)}-\epsilon_{\alpha}^{(n)})^2+\gamma_\alpha^2
        },\\
        =&
        \sum_{m,n}
        \frac{1}{\pi}\frac{
            \gamma_\alpha
        }{
            (\mu_F^{(m)}-\epsilon_{\alpha}^{(n)})^2+\gamma_{\alpha}^2
        }
        \frac{\left\langle
            \phi_{\alpha}^{(n)}
            \left|
            \pi \Gamma_L
            \right|
            \phi_{\alpha}^{(n)}
            \right\rangle}{\gamma_{\alpha}}\label{eq:division}
    \end{align}  
    where the second line in Eq.~\eqref{eq:division} is obtain by dividing and multiplying $\gamma_{\alpha}$.
    Furthermore, let us define
   \begin{equation}\label{eq:def_w_n1}
    w_{\alpha}^{(n)} =
    \frac{\left\langle
            \phi_{\alpha}^{(n)}
            \left|
            \pi \Gamma_L
            \right|
            \phi_{\alpha}^{(n)}
            \right\rangle}{\gamma_{\alpha}}.
    \end{equation}
   Using the definition of the weight function in Eq.~\eqref{eq:def_w_n1}, Eq.~\eqref{eq:division} can be recasted as      

    \begin{equation}
    \mathcal{\bar{F}}^{(0)}_{k_y,\alpha\alpha}(\mu_F)
		=
		\sum_{m,n}
		\frac{1}{\pi}\frac{\gamma_{\alpha}}
        {(\mu_F^{(m)} - \epsilon_\alpha - n\Omega)^2 + \gamma_\alpha^2}\,w_{\alpha}^{(n)}.\label{eq:Fdiag_simple}
    \end{equation}
    
    \noindent Furthermore in the weak-coupling limit $\gamma\rightarrow 0$, we have the following relation
    \begin{align}
    \lim_{\gamma\rightarrow 0}
    \frac{1}{\pi}\frac{\gamma_{\alpha}}
    {(\mu_F^{(m)} - \epsilon_\alpha - n\Omega)^2 + \gamma_\alpha^2}
    =
    \delta\big(\mu_F^{(m)} - \epsilon_{\alpha}^{(n)}\big).\label{eq:delta_wc}
    \end{align}
    Substituting Eq.~\eqref{eq:delta_wc} into Eq.~\eqref{eq:Fdiag_simple} leads to
    \begin{equation}
        \mathcal{\bar{F}}^{(0)}_{k_y,\alpha\alpha}(\mu_F)
		=
		\sum_{m,n}
		\delta\big(\mu_F^{(m)} - \epsilon_{\alpha}^{(n)}\big)\,w_{\alpha}^{(n)}.\label{eq:Fdiag_simple2}
    \end{equation}
      
   	By substituting Eq.~\eqref{eq:Jdiag_simple} and \eqref{eq:Fdiag_simple2} into Eq.~\eqref{eq:Ghall-diag_q0-1}, and converting the $k_y$ summation to an integral, i.e. $\frac{2\pi}{n_y}\sum_{k_y}\rightarrow\int_0^{2\pi} dk_y$ we obtain the following 
    \begin{align}
        \overline{\mathcal{G}}^{\text{avg.}}_H(\mu_F)
    	=&
    	\sum_{\alpha,n,m} \int_0^{2\pi} dk_y 
    	w_{\alpha}^{(n)} 
    	\delta(\mu_F^{(m)} - \epsilon_\alpha^{(n)}) 
    	\frac{d\epsilon_\alpha}{dk_y}.
    	\label{eq:Gy-weak-n}
    \end{align}
    We remark that Eq.~\eqref{eq:Gy-weak-n} can also be derived from the steady-state Floquet density matrix in the weak-coupling limit (see Appendix.~\ref{app:EZF} for details).
     Furthermore, since both $\mu_F$ and $\epsilon_\alpha$ have values in the range $(-\Omega/2,\Omega/2)$, the above sum gets contributions only when $m=n$. We then have the simplification
    \begin{align}
     &\overline{\mathcal{G}}^{\text{avg.}}_H(\mu_F)  
     =
    	\sum_{\alpha,n} \int_0^{2\pi} dk_y 
    	w_{\alpha}^{(n)} 
    	\delta(\mu_F - \epsilon_\alpha) 
    	\frac{d\epsilon_\alpha}{dk_y} \\
     =&\sum_{\alpha} \int_0^{2\pi} dk_y 
    	\left(\sum_n w_{\alpha}^{(n)} \right)
    	\delta(k_y-k_y^*) 
    	\frac{d\epsilon_\alpha/dk_y}{|d\epsilon_\alpha/dk_y|}, \\
        &=
    	\sum_{\alpha}
    	\left(\sum_{n}w_{\alpha}^{(n)}\right) \operatorname{sgn}\!\left(\partial_{k_y}\epsilon_\alpha\right)\Big|_{k_y*},
        \label{eq:gavga}
    \end{align}
    where $k_y^*$ is determined from the condition $\mu_F = \epsilon_{\alpha}(k_y)$. 
   Next we look at the  weight function $ w_{\alpha}^{(n)}$ and examine the contributions from the bulk bands and the edge modes  localized at the left and right boundaries of the cylinder.
   Using the expression of $\gamma_{\alpha}$ in Eq.~\eqref{eq:expan} we get from  Eq.~\eqref{eq:def_w_n1} 
     \begin{align}
        w_{\alpha}^{(n)} =
        \frac{\left\langle
                \phi_{\alpha}^{(n)}
                \left|
                \pi \Gamma_L
                \right|
                \phi_{\alpha}^{(n)}
                \right\rangle}{\sum_{p}\left\langle
                \phi_{\alpha}^{(p)}
                \left|
                \pi \Gamma
                \right|
                \phi_{\alpha}^{(p)}
                \right\rangle}%\label{eq:def_w_n2}
                =
               \frac{\left|\phi_{\alpha}^{(n)}(x_{\mathrm{L}})\right|^2}{\sum_{p,\lambda} \left|\phi_{\alpha}^{(p)}(x_{\lambda})\right|^2},\label{eq:def_w_n3}
    \end{align}
where $x_L=1$ and $x_R=2n_x$.  The above form follows from the structure $\Gamma_{L,x,x'}=\gamma \delta_{xx'}\delta_{x,1}$ and $\Gamma_{R,x,x'}=\gamma \delta_{xx'}\delta_{x,2 n_x}$. We now make the following observations:

\begin{enumerate}
    \item First consider the case when $\mu_F$ is in a topological gap. In this case only the two chiral modes, localized in the left and right ends, contribute to the $\alpha$ sum. For the  left mode,  $|\phi_{\alpha}^{(n)}(x_{\mathrm{L}})|^2$ is significant while  
$|\phi_{\alpha}^{(n)}(x_{\mathrm{R}})|^2$
is small and vice-versa for the right edge mode. Hence $w_\alpha^{(n)}$ is negligible for the right edge mode. For the left mode the contribution of the right mode in the denominator can be neglected and we get:
\begin{align}
    w_{\alpha}^{(n)} =
    \frac{\left|\phi_{\alpha}^{(n)}(x_{\mathrm{L}})\right|^2}{\sum_{p} \left|\phi_{\alpha}^{(p)}(x_L)\right|^2}.
    \label{eq:def_w_f} 
\end{align}
Summing over all side bands ($n$) and using Eq.~\eqref{eq:gavga} we finally get our required result
\begin{align}
     &\overline{\mathcal{G}}^{\text{avg.}}_H(\mu_F)  
     =\pm 1,
     \end{align}
where the sign is determined by that of the group velocity of the left edge mode $\partial \epsilon(k_y)/\partial k_y$ at $\mu_F$.  The above equation captures the net chirality of Floquet edge modes at quasienergy $\mu_F$, i.e., the Floquet winding invariant~\cite{Rudner2013}. Hence, we can make the identification: 
    \begin{equation}
    	\overline{\mathcal{G}}^{\text{avg.}}_H(\mu_F) = W_{\mu_F}.
    \end{equation}

\item We note that in addition to the quantization of the summed conductance our weak-coupling approach also allows us to compute the contribution of each side-band to the total conductance and this is given by Eq.~\eqref{eq:def_w_f}.

\item  Next we consider the case where $\mu_F$ lies in the bulk bands. In this case the weights, $|\phi_\alpha(X_\lambda)|^2$,
at both left and right ends are small but comparable and so we get $w_\alpha^{(n)} <1$. Secondly there are a large number of $\alpha$ modes which will contribute to the sum and hence we do not get any quantization.
\end{enumerate}

    This completes our proof of the Floquet sum rule and quantization of the Hall conductance in the weak coupling limit. 

    \section{Summary and Outlook}	
    \label{sec:Summary and Outlook}
    In this work we considered
     a simple tight-binding model on a square lattice where the edge hoppings are switched off an on in a specified time-periodic way. This Floquet system is known to support chiral edge modes
    and the topology of the Floquet bands is characterized by the Floquet winding invariants, as periodic driving can lead to anomalous phases in which the system is topological even when all bulk Chern numbers are zero. 
    We studied the open system set up using the formalism of Floquet NEGF and  demonstrated numerically the quantization of the two-terminal (longitudinal) and Hall (transverse)  conductance, showing that these transport properties directly captures the underlying  Floquet winding number, including its magnitude and sign. Our results thus establish a practical route to measure these winding  invariants through longitudinal and transverse current measurements in two- and four-terminal setups, providing a direct correspondence between measurable transport and topology. We also evaluated local currents inside the sample, which provide a more detailed characterization of transport.  In addition the local bond currents along the  $y$ direction enables us to compute the Hall conductance in the strip geometry. 
    
    Numerically, we found that perfect quantization emerges in the weak-coupling regime for sufficiently large systems.  The saturation of the conductance quantization has a weaker dependence on $n_x$, while a stronger dependence on $n_y$. Using the weak-coupling limit  in a cylindrical geometry, we then  provide an analytical proof of Hall conductance quantization.
    As a result of periodic driving, the net conductance gets distributed across the Floquet sidebands. The microscopic weak-coupling  calculcation    gives explicitly the contribution of each sideband to the total conductance and shows that adding their contributions restores quantazation, thus proving the  
       Floquet sum rule for the Hall conductance.

    In this weak coupling limit, transport is governed by intraband processes that can be directly related to the eigenspectrum of the isolated Hermitian Floquet system. Tuning the chemical potential selects contributions from individual chiral modes in different sidebands. Our results show that the contributions of a few bands is sufficient to give us near perfect quantization, hence these could be accessible to experiments. 
    
    These results provide a unified framework connecting transport quantization and Floquet band topology, thereby clarifying the microscopic origin of quantized Hall response in driven open systems. Future directions include exploring strong-coupling and non-perturbative regimes, where interband processes are expected to become relevant, as well as investigating transient current dynamics during the approach to the steady state.

	\section{Acknowledgements}
    We thank S.~Rao, G.~Murthy, B.~Dou\c{c}ot, R.~Shankar, S.~Roy and D.~Sen for valuable discussions. A.D. acknowledges support from the J.C.~Bose Fellowship (JCB/2022/000014) of the Science and Engineering Research Board, Department of Science and Technology, Government of India. We also acknowledge support from the Department of Atomic Energy, Government of India, under Project No. RTI4001. R.K. and M.K. acknowledge the LOTUS Programme by the Japan Science and Technology Agency (JST) and the Advanced Institute for Materials Research, Tohoku University, Sendai, Japan.  
    
    %\clearpage
	\setcounter{figure}{0}
	\renewcommand{\thefigure}{A\arabic{figure}}
	\appendix\section{Glossary of symbols}
	\label{sec:Glossary of symbols}
	\begin{table}[ht!]
		\centering
		{\footnotesize
			\setlength{\tabcolsep}{4pt}
			\renewcommand{\arraystretch}{0.95}
			\begin{tabular}{|c|p{0.6\columnwidth}|}\hline
				\textbf{Symbol} & \textbf{Meaning} \\\hline
				$h(\mathbf{k},t)$ & Bloch Hamiltonian \\ \hline
				$U(\mathbf{k},t)$ & Time-evolution operator \\ \hline
				$U(\mathbf{k},T)$ & One-period time-evolution operator \\ \hline
				$\epsilon_\alpha(\mathbf k)$ & Quasienergy of the closed system \\ \hline
				$\epsilon_\alpha^{(n)}(\mathbf k)$ &  Quasienergy in $n$th Floquet replica \\ \hline
				$|\phi_\alpha(\mathbf k, t)\rangle$ & Floquet mode of the closed system\\ \hline
				$|\phi_\alpha^{(n)}(\mathbf k)\rangle$ & $n$th Floquet harmonic of the Floquet mode of the closed system \\ \hline
				$C_\alpha$ & Chern number of band $\alpha$ \\ \hline
				$W_0, W_\pi$ & Floquet winding numbers \\ \hline
				$C_+$ & Chern number of the upper band \\ \hline
				%$\rho_\lambda$ & Density of states of lead $\lambda$ \\ \hline
				$t_{\lambda}$ & Tunneling amplitudes between system and $\lambda$th reservoir \\ \hline
                $\gamma$ & System-reservoir coupling parameter \\ \hline
				$G(t,t')$ & Retarded Green’s function \\ \hline
				$G^{(q)}(\omega)$ & $q$th Floquet harmonic of the Green’s function \\ \hline
				$T^{(q)}_{\lambda\lambda'}$ & Photon-assisted transmission coefficient \\ \hline
				$I_\lambda(\mu_F)$ & Current in lead $\lambda$ \\ \hline
				$G_2(\mu_F)$ & Two-terminal conductance \\ \hline
				$\overline{G}_2(\mu_F)$ & Summed two-terminal conductance \\ \hline
				$J_{r\to r'}$ & Bond current from site $r$ to $r'$ \\ \hline
				$\mathcal G_x(\mu_F), \mathcal G_y(\mu_F)$ & Longitudinal and transverse conductances\\ \hline
				$\overline{\mathcal{G}}_{2,x}(\mu_F)$ & Summed longitudinal conductance \\ \hline
				$\overline{\mathcal{G}}_{H,y}(\mu_F)$ & Summed Hall conductance \\ \hline
                $e + i\gamma_{\alpha}$ & Complex quasienergy eigenvalue of the open system \\ \hline
                $|u_\alpha^{+}\rangle, |u_\alpha^{-}\rangle$ & Biorthogonal Floquet eigenmodes of open system \\ \hline
				$\mathcal{F}^{(q)}_{\alpha\beta}$ & $q$th Floquet harmonic of cross-occupation functions \\ \hline
				$J^{(q)}_{k_y,\alpha\beta}$ & $q$th harmonic of the current matrix element \\ \hline
				$w_\alpha^{(n)}$ & Spectral weight of the $n$th Floquet harmonic at the left junction \\ \hline
				$n_x, n_y$ & Number of Bravais lattice sites along $x$ and $y$ (strip geometry) \\ \hline
                $n_x, n_y/2$ & Number of Bravais lattice sites along $x$ and $y$ directions (cylinder geometry) \\ \hline
			\end{tabular}
		}
        \caption{List of some of the notations used throughout the main text and appendices.}
        \label{app:table}
	\end{table}
    This appendix collects (see Table.~\ref{app:table}) some of the notations used throughout the main text and appendices. It is intended to provide a quick reference for the symbols appearing in the Floquet Hamiltonian, Green’s functions, and transport expressions introduced in Secs.~\ref{sec:Floquet topological phases}--\ref{sec:Numerical results}.    
    %Here, $\gamma_\alpha$ is obtained from the matrix elements of $\Gamma = \Gamma_L + \Gamma_R$, evaluated in the Floquet eigenbasis. T
	
	\section{Floquet Extended Zone Formalism}\label{app:EZF}
	This appendix provides the technical details of the Floquet--Sambe (extended-zone) formalism used to compute quasienergies and Floquet eigenstates in Sec.~\ref{sec:Floquet topological phases}. In particular, it includes the calculation of the quasienergy spectra shown in Fig.~\ref{fig:FIG2_QES}, as well as the definitions of Floquet harmonics employed throughout the transport analysis.
	
	We consider a periodically driven system described by a single-particle Bloch Hamiltonian $h(\mathbf{k},t)$, whose dynamics are governed by the time-dependent Schr\"odinger equation
	\begin{equation}
		i\partial_t |\psi_{\alpha}(\mathbf{k},t)\rangle
		=
		h(\mathbf{k},t)|\psi_{\alpha}(\mathbf{k},t)\rangle ,
		\label{eq:tdshe-supp}
	\end{equation}
	where $h(\mathbf{k},t+T)=h(\mathbf{k},t)$ with $T=2\pi/\Omega$ the driving period. Throughout this appendix we set $e=1$ and $\hbar=1$. According to Floquet theory, the solutions of Eq.~\eqref{eq:tdshe-supp} take the form
	\begin{equation}
		|\psi_{\alpha}(\mathbf{k},t)\rangle
		=
		e^{-i\epsilon_{\alpha}(\mathbf{k}) t}
		|\phi_{\alpha}(\mathbf{k},t)\rangle ,
		\label{eq:ualpha-supp}
	\end{equation}
	where $|\phi_{\alpha}(\mathbf{k},t)\rangle$ are time-periodic functions with period $T$, known as Floquet modes, and $\epsilon_{\alpha}(\mathbf{k})$ are the corresponding quasienergies. The index $\alpha$ labels distinct Floquet bands and runs over the dimension of the physical Hilbert space associated with $h(\mathbf{k},t)$.
	
	Substituting Eq.~\eqref{eq:ualpha-supp} into Eq.~\eqref{eq:tdshe-supp} yields the Floquet eigenvalue equation
	\begin{equation}
		\left[h(\mathbf{k},t)-i\partial_t\right]
		|\phi_{\alpha}(\mathbf{k},t)\rangle
		=
		\epsilon_{\alpha}(\mathbf{k})
		|\phi_{\alpha}(\mathbf{k},t)\rangle .
		\label{eq:SHE-ualpha}
	\end{equation}
	
	\noindent To solve Eq.~\eqref{eq:SHE-ualpha}, we expand both the Hamiltonian and the Floquet modes in a Fourier series,
	\begin{eqnarray}
		h(\mathbf{k},t)
		&=&
		\sum_{m\in\mathbb{Z}} e^{-im\Omega t}\, h^{(m)}(\mathbf{k}),\\
		|\phi_{\alpha}(\mathbf{k},t)\rangle
		&=&
		\sum_{n\in\mathbb{Z}} e^{-in\Omega t}\,
		|\phi_{\alpha}^{(n)}(\mathbf{k})\rangle ,
		\label{eq:FT_ut}
	\end{eqnarray}
	where $m,n$ label Floquet (extended-zone) harmonics. Inserting these expansions into 
	Eq.~\eqref{eq:SHE-ualpha} and equating coefficients of
	$e^{-im\Omega t}$ yields the following extended-zone (Floquet--Sambe) eigenvalue equation for each $m$
	\begin{equation}
		\sum_{n\in\mathbb{Z}}
		\left(h^{(m-n)}(\mathbf{k})-\delta_{m,n}n\Omega\mathbb{I}\right)
		|\phi_{\alpha}^{(n)}(\mathbf{k})\rangle
		=
		\epsilon_{\alpha}(\mathbf{k})
		|\phi_{\alpha}^{(m)}(\mathbf{k})\rangle ,
		\label{eq:EZF_derivative}
	\end{equation}
	The resulting Floquet--Sambe Hamiltonian, given in Eq.~\eqref{eq:EZF_derivative}, has a block structure with diagonal matrix elements (of size $2\times 2$) shifted by $h^{(0)}-n\Omega\mathbb{I}$ and off-diagonal blocks given by $h^{(m-n)}$. Its eigenvalues are $\epsilon_{\alpha}(\mathbf{k}) + n\Omega$, which define the quasienergies and their associated sidebands. The quasienergies $\epsilon_{\alpha}(\mathbf{k})$ are defined modulo $\Omega$, i.e. $ \Omega/2\leq\epsilon_{\alpha}(\mathbf{k})\leq\Omega/2$..
    
    The dimension of the extended (Floquet–Sambe) space is given by the tensor product of the physical Hilbert space and the Fourier space of T-periodic functions. For example, if the dimension of the physical space for the strip geometry is \(2 n_x n_y\), and the dimension of the Fourier space of the time-periodic functions is \(2N_F + 1\), then the dimension of the Floquet--Sambe space is given by \((2N_F + 1)\times N\). In practical calculations, the infinite Floquet–Sambe Hamiltonian is truncated to a finite number of harmonics, $|n|\leq N_F$.
	
	\noindent The Fourier components of normalized Floquet modes $|\phi_{\alpha}^{(n)}(\mathbf{k})\rangle$ satisfy the orthonormality and
	completeness relation, i.e.
	\begin{eqnarray}
		\sum_{m}
		\langle \phi_{\alpha}^{(m)}(\mathbf{k})
		| \phi_{\beta}^{(m)}(\mathbf{k}) \rangle
		&=&
		\delta_{\alpha\beta},\\
		\sum_{\alpha,m}
		|\phi_{\alpha}^{(m)}(\mathbf{k})\rangle
		\langle \phi_{\alpha}^{(m)}(\mathbf{k})|
		&=&
		\mathbb{I},
	\end{eqnarray}
	where $\mathbb{I}$ denotes the identity matrix on the physical Hilbert space.
	An expression for the velocity of the quasienergy bands can be obtained by differentiating Eq.~\eqref{eq:EZF_derivative} with respect to $\mathbf{k}$, yielding
	\begin{align}
		&\sum_{n}
		\partial_{\mathbf{k}} h^{(m-n)}(\mathbf{k})\,
		|\phi_{\alpha}^{(n)}(\mathbf{k})\rangle
		+
		\sum_{n}
		\left[h^{(m-n)}(\mathbf{k})-\delta_{m,n}n\Omega\mathbb{I}\right]
		\nonumber\\
		&\partial_{\mathbf{k}}
		|\phi_{\alpha}^{(n)}(\mathbf{k})\rangle=
		\partial_{\mathbf{k}}\epsilon_{\alpha}(\mathbf{k})\,
		|\phi_{\alpha}^{(m)}(\mathbf{k})\rangle
		+
		\epsilon_{\alpha}(\mathbf{k})\,
		\partial_{\mathbf{k}}
		|\phi_{\alpha}^{(m)}(\mathbf{k})\rangle.
	\end{align}
	By rearranging terms in the above equation, we obtain the following for each $m$
	\begin{align}
		&
		\partial_{\mathbf{k}}\epsilon_{\alpha}(\mathbf{k})\,
		|\phi_{\alpha}^{(m)}(\mathbf{k})\rangle
		-
		\sum_{n}
		\partial_{\mathbf{k}} h^{(m-n)}(\mathbf{k})\,
		|\phi_{\alpha}^{(n)}(\mathbf{k})\rangle
		\nonumber\\
		&=
		\sum_{n}
		\Big[
		h^{(m-n)}(\mathbf{k})
		-\delta_{m,n}n\Omega\mathbb{I}
		-\delta_{m,n}\epsilon_{\alpha}(\mathbf{k})
		\Big]
		\partial_{\mathbf{k}}
		|\phi_{\alpha}^{(n)}(\mathbf{k})\rangle.\label{eq:FHth_1}
	\end{align}
	\noindent Defining the operator
	\begin{equation}
		\mathcal{O}^{(m,n)}_{\alpha}
		=
		h^{(m-n)}(\mathbf{k})
		-\delta_{m,n}\left(n\Omega
		+\epsilon_{\alpha}(\mathbf{k})\right)\mathbb{I}\,,\label{eq:def_Omn}
	\end{equation}
    and substituting Eq.~\eqref{eq:def_Omn} into Eq.~\eqref{eq:FHth_1}, we obtain
	\begin{align}\nonumber
		\partial_{\mathbf{k}}\epsilon_{\alpha}(\mathbf{k})\,
		|\phi_{\alpha}^{(m)}(\mathbf{k})\rangle
		&=
		\sum_{n}
		\partial_{\mathbf{k}} h^{(m-n)}(\mathbf{k})\,
		|\phi_{\alpha}^{(n)}(\mathbf{k})\rangle 
		\\
		&+
		\sum_{n}
		\mathcal{O}^{(m,n)}_{\alpha}
		\partial_{\mathbf{k}}
		|\phi_{\alpha}^{(n)}(\mathbf{k})\rangle .
	\end{align}
	
	\noindent Projecting onto $\langle \phi_{\alpha}^{(m)}(\mathbf{k})|$ and summing over $m$, we obtain
	\begin{eqnarray}
		\partial_{\mathbf{k}}\epsilon_{\alpha}(\mathbf{k})
		&=&
		\sum_{m,n}
		\langle \phi_{\alpha}^{(m)}(\mathbf{k}) |
		\partial_{\mathbf{k}} h^{(m-n)}(\mathbf{k})
		| \phi_{\alpha}^{(n)}(\mathbf{k}) \rangle
		\nonumber\\
		&&+
		\sum_{m,n}
		\langle \phi_{\alpha}^{(m)}(\mathbf{k}) |
		\mathcal{O}^{(m,n)}_{\alpha}
		\partial_{\mathbf{k}}
		|\phi_{\alpha}^{(n)}(\mathbf{k})\rangle.\label{eq:FHt_2}
	\end{eqnarray}
	Substituting the Floquet eigenvalue equation [Eq.~\eqref{eq:EZF_derivative}], we obtain
	\begin{equation}
	\sum_{m}
	\langle \phi_{\alpha}^{(m)}(\mathbf{k}) |
	\mathcal{O}^{(m,n)}_{\alpha}
	= 0,
	\end{equation}
    and substituting the above into Eq.~\eqref{eq:FHt_2} yields the following:	
    \begin{equation}
		\partial_{\mathbf{k}}\epsilon_{\alpha}(\mathbf{k})
		=
		\sum_{m,n}
		\langle \phi_{\alpha}^{(m)}(\mathbf{k}) |
		\partial_{\mathbf{k}} h^{(m-n)}(\mathbf{k})
		| \phi_{\alpha}^{(n)}(\mathbf{k}) \rangle .
		\label{eq:FloquetHF_final}
	\end{equation}
	Eq.~\eqref{eq:FloquetHF_final} is the Floquet Hellmann--Feynman theorem~\cite{FFHT_Sambe_ref1,FFHT_Giuseppe_ref2,FFHT_Qian_ref3,FFHT_shtoff_ref4}, which expresses the velocity of quasienergy bands entirely in terms of the Fourier components of the Hamiltonian and the Floquet eigenstates of a closed system. This relation is used in the main text to evaluate expectation of the current operator in terms of the quasienergy band velocity [Eq.~\eqref{eq:Jdiag_simple}] and Hall conductance [Eq.~\eqref{eq:Gy-weak-n}].

    Additionally, we would also like to highlight here, that the Hall conductance expression in the weak coupling limit [Eq.~\eqref{eq:Gy-weak-n}] can also be expressed in terms of the steady state density matrix in weak coupling limit. For a cylindrical setup connected to two reservoirs, as shown in Fig.~\ref{fig:cylindrical_geo}, the steady-state single-particle density matrix of the driven central region, obtained after integrating out the reservoirs, can be written as (see Ref.~\cite{Kumari2024})
	\begin{equation}
    \rho(k_y,t)=\sum_{\alpha}n_\alpha(k_y)|\phi_\alpha(k_y,t)\rangle\langle\phi_\alpha(k_y,t)|,\label{eq:dmat_SS}
	\end{equation}
	where $|\phi_\alpha(k_y,t)\rangle$ are the Floquet modes of the driven system. The occupations that appear in Eq.~\eqref{eq:dmat_SS}, are given by~\cite{Kumari2024},
	\begin{equation}
		n_\alpha(k_y)=\gamma_\alpha(k_y)^{-1}\sum_{\lambda,n}
		\langle \phi^{(n)}_\alpha(k_y)|\Gamma_\lambda|\phi^{(n)}_\alpha(k_y)\rangle 
		f_{\lambda,\alpha}^{(n)},
	\end{equation}
	where $\gamma_\alpha(k_y)=\sum_{\lambda,l}\la \phi_{\alpha}^{(l)}(k_y)|\Gamma_{\lambda}|\phi_{\alpha}^{(l)}(k_y)\ra$, $f_{\lambda,\alpha}^{(n)}=f(\epsilon_\alpha(k_y)+n\Omega-\mu_\lambda)$ and $k_y$ is the momentum along the periodic direction. We recall that $\lambda=L, R$ corresponds to the left, right reservoir and $l$ labels the Floquet extended zone harmonics. For a small bias $\Delta\mu$ between the reservoirs, with $\mu_L=\mu_F+\Delta\mu$ and $\mu_R=\mu_F$ the excess occupation in linear response, that is responsible for the net current flow, is given by
    \begin{equation}
		\Delta n_\alpha(k_y,\mu_F)= -\frac{\Delta\mu}{\gamma_\alpha}
		\sum_n\langle \phi^{(n)}_\alpha|\Gamma_L|\phi^{(n)}_\alpha\rangle 
		\left.\frac{\partial f(\epsilon^{(n)})}{\partial \epsilon}\right|_{\epsilon^{(n)}=\mu_F}.\label{eq:excess_occu}
	\end{equation}
	For brevity, we henceforth suppress the $k_y$ and $\mu_F$ dependence in Eq.~\eqref{eq:excess_occu}. The time-averaged excess Hall current follows from
	\begin{align}
		\mathcal{J}_y(\mu_F)
		&=\frac{1}{T}\int_0^T dt \int_0^{2\pi} \frac{dk_y}{2\pi}\,
		\mathrm{Tr}\!\left[\Delta\rho(t)J_y\right]\,,
        %\\
		%&=\sum_{\alpha}\int_0^{2\pi} \frac{dk_y}{2\pi}\,\Delta n_\alpha\,\mathcal{J}^{(0)}_{k_y,\alpha\alpha},
		\label{eq:excess-jy}
	\end{align}
    where $J_y(k_y)=\frac{\partial h(k_y,t)}{\partial k_y}$ is the current operator 
    %given by $\frac{\partial h(k_y,t)}{\partial k_y}$, 
    and 
$\Delta\rho(t)=\sum_{\alpha}\Delta n_\alpha(k_y,\mu_F)|\phi_\alpha(k_y,t)\rangle\langle\phi_\alpha(k_y,t)|$, where $\Delta n_\alpha(k_y,\mu_F)$ is defined in Eq.~\eqref{eq:excess_occu}. Furthermore, in the weak-coupling limit, the occupations $\Delta n_\alpha(k_y)$ become time-independent~\cite{Kumari2024}, and we obtain
    \begin{align}\nonumber
         \mathcal{J}_y(\mu_F)&=\int_0^{2\pi}\frac{dk_y}{2\pi}\Delta n_\alpha\int_0^{T}
         \frac{dt}{T}\langle\phi_\alpha(t)|\frac{\partial h(t)}{\partial k_y}|\phi_\alpha(t)\rangle \\   \nonumber 
        &=\int_0^{2\pi} \frac{dk_y}{2\pi} \Delta n_\alpha\sum_{m,n}\la\phi^{(m)}|\partial_{k_y}h^{(m-n)}|\phi^{(n)}\ra\\    
        &=\int_0^{2\pi} \frac{dk_y}{2\pi} \Delta n_\alpha\partial_{k_y} \epsilon_\alpha ,\label{eq:J_density_matrix}
        \end{align}
    where in order to get the last line in Eq.~\eqref{eq:J_density_matrix}, we used the Floquet Hellmann--Feynman relation given in Eq.~\eqref{eq:FloquetHF_final}.
     Next, we define the summed Hall conductance, by using the Floquet sum rules, i.e., $\overline{\mathcal{G}}^{\text{avg.}}_{H}(\mu_F)=2\pi\sum_m\mathcal{J}_y(\mu_F+m\Omega)/\Delta\mu$ (in units of $e^2/h$), which yields [using Eq.~\eqref{eq:J_density_matrix}]
	\begin{equation}
		\overline{\mathcal{G}}^{\text{avg.}}_{H}(\mu_F)=\sum_{\alpha,n}\int_0^{2\pi} dk_y\,
		w_{\alpha}^{(n)}
		\frac{\partial\epsilon_\alpha}{\partial k_y}
		\delta(\epsilon^{(n)}_\alpha-\mu_F^{(m)}),
		\label{eq:G_h_density}
	\end{equation}
	where   $w_{\alpha}^{(n)} =
    \frac{\left\langle
    \phi_{\alpha}^{(n)}
    \left|\Gamma_L\right|\phi_{\alpha}^{(n)}
    \right\rangle}{\sum_{p}\left\langle\phi_{\alpha}^{(p)}
    \left|\Gamma\right|\phi_{\alpha}^{(p)}
    \right\rangle}$. Eq.~\eqref{eq:G_h_density} is identical to Eq.~\eqref{eq:Gy-weak-n} of the main text. 
    
	\section{Floquet winding invariants}
	\label{sec: Floquet winding invariants}
	In this appendix, we review the definition of Floquet winding invariants following Refs.~\cite{Rudner2013,Unal2019PRR}. These invariants provide the bulk topological classification of two-dimensional periodically driven systems and determine the quantized transport responses discussed in Sec.~\ref{sec:Numerical results}.
	
	We consider a two-dimensional periodically driven Bloch Hamiltonian given by $h(\mathbf{k},t)=h(\mathbf{k},t+T)$, where $\mathbf{k}=(k_x,k_y)$ is the crystal momentum. The corresponding time-evolution operator is $U(\mathbf{k},t)=\mathcal{T}\exp\!\left[-i\int_0^t h(\mathbf{k},t')\,dt'\right]$, which is unitary for all $t$. For a fixed $\mathbf{k}$, the evolution over one driving period defines the Floquet operator $U(\mathbf{k},T)$, whose eigenvalues ($\epsilon_{\alpha}(\mathbf{k})$) determine the quasienergy spectrum modulo $2\pi/T$, i.e. $-\Omega/2\leq\epsilon_{\alpha}(\mathbf{k})\leq\Omega/2$.
	
	As a function of $(k_x,k_y,t)$, the evolution operator defines a continuous mapping
	\begin{equation}
		U(\mathbf{k},t): \mathbb{T}^3 \equiv (k_x,k_y,t) \longrightarrow U(N),
	\end{equation}
	where $N$ is the number of internal degrees of freedom (bands, which is $2$ in our case) of the Bloch Hamiltonian. If the Floquet operator satisfies $U(\mathbf{k},T)=\mathbb{I}$, where $\mathbb{I}$ is the identity matrix of $N\times N$ dimensions, the evolution is strictly periodic over one driving cycle and defines a closed map on the three-dimensional torus $\mathbb{T}^3$. Such a map is classified by an integer-valued winding number
	\begin{equation}
		W[U]=\frac{1}{8\pi^2}\int d^2k\,dt\;
		\mathrm{Tr}\!\left(
		U^{-1}\partial_t U
		\bigl[
		U^{-1}\partial_{k_x}U,
		U^{-1}\partial_{k_y}U
		\bigr]
		\right),
		\label{eq:winding_def}
	\end{equation}
    which is a topological invariant that remains unchanged under any smooth deformation in the evolution that does not close the quasienergy gap.
    
	In general, however, the physical Floquet operator does not satisfy $U(\mathbf{k}, T)=\mathbb{I}$, and the time evolution does not define a closed map on $\mathbb T^3$. To construct a topological invariant in this case, one introduces a modified evolution operator $U_\varepsilon(\mathbf{k},t)$ that restores periodicity by smoothly connecting the physical evolution to a contractible evolution generated by an effective Floquet Hamiltonian. Following Ref.~\cite{Rudner2013}, this modified evolution is defined as
	\begin{equation}
		U_\varepsilon(\mathbf{k},t)=
		\begin{cases}
			U(\mathbf{k},2t), & 0\le t\le T/2, \\[4pt]
			V_\varepsilon(\mathbf{k},2T-2t), & T/2\le t\le T,
		\end{cases}
		\label{eq:Ueps_def}
	\end{equation}
	where
	\begin{equation}
		V_\varepsilon(\mathbf{k},t)=\exp[-iH_{\mathrm{eff}}(\mathbf{k})t], \qquad
		H_{\mathrm{eff}}(\mathbf{k})=\frac{i}{T}\log_{\varepsilon} U(\mathbf{k},T),
		\label{eq:Heff_def}
	\end{equation}
    where the symbol $\log_{\varepsilon}$ in Eq.~\eqref{eq:Heff_def}, indicates that the branch-cut in logarithmic is chosen at $\varepsilon$, corresponding to the selected quasienergy gap. The modified evolution in Eq.~\eqref{eq:Ueps_def}, then satisfies $U_\varepsilon(\mathbf{k},T)=\mathbb{I}$ by construction, and thus defines a closed map on $T^3$.
	
	The winding number $W_\varepsilon$, evaluated for $U_\varepsilon$, is the Floquet winding invariant $W[U_\varepsilon]$, associated with the quasi-energy gap centered at $\varepsilon$. This expression $W[U_\varepsilon]$ is well-defined provided the evolution operator is smooth and the quasienergy gaps remain open throughout the cycle. In particular, the choices $\varepsilon = 0$ and $\varepsilon = \pi/T$ define two independent invariants, $W_0$ and $W_\pi$, characterizing the topology of the zero and $\pi$ quasienergy gaps, respectively. These invariants are related to the Chern number $C_+$ of the Floquet bands below the $\pi$ gap via~\cite{Rudner2013},
	\begin{equation}
		W_\pi - W_0 = C_+.
		\label{eq:winding_chern_relation}
	\end{equation}
	This relation emphasizes that, in contrast to static systems, Chern numbers alone are insufficient to fully characterize the topology of periodically driven systems.
    
	For systems with open boundaries, the bulk–boundary correspondence is encoded in the winding invariants: the number of chiral edge modes traversing the quasienergy gap at $\varepsilon$ equals $W_\varepsilon$. The Floquet winding invariants thus provide a complete topological characterization of two-dimensional periodically driven systems, capturing both conventional and anomalous Floquet edge states.
	
	\section{Floquet--NEGF}\label{sec:Floquet_NEGF}
    This appendix presents the detailed derivation of the Floquet nonequilibrium Green's-function (NEGF) formalism used in Sec.~\ref{sec:NEGF_formalism_for_Floquet_systems}. 
    
    We briefly review the Floquet nonequilibrium Green’s-function (NEGF) formalism, which provides a general framework for transport in periodically driven open quantum systems~\cite{Kohler2005}. We consider a driven central region coupled to two macroscopic, static (non-driven) fermionic reservoirs in the wide-band limit, as illustrated in Fig.~\ref{fig:FIG-SETUP} of the main text. The left and right reservoirs are labeled by $\lambda = L$ and $R$, respectively.
	
    The tight-binding Hamiltonian of the full system is given by
    \begin{equation}\label{eq:Hfulla-app}
    \mathcal{H}(t)
    = H(t) + \sum_{\lambda=L,R} \left( H_{S\lambda} + H_\lambda \right),
    \end{equation}
    where $H_\lambda$ describes the $\lambda$-th reservoir, $H_{S\lambda}$ denotes the coupling between the system and the $\lambda$-th reservoir, and $H(t)$ is the time-dependent Hamiltonian of the driven central region, defined on a finite two-dimensional lattice [see Eq.~\eqref{eq:h5step_strip}].  Each site in the full system is labeled by $r=(i,j)$, where $i = 1, 2, \ldots, n_L + n_R + 2n_x$ and $j = 1, 2, \ldots, n_y$. The left reservoir occupies the region $i = 1, \ldots, n_L$ and $j = 1, 2, \ldots, n_y$, the driven system occupies $i = n_L + 1, \ldots, n_L + 2n_x$ and $j = 1, 2, \ldots, n_y$, and the right reservoir occupies $i = n_L + 2n_x + 1, \ldots, n_L + n_R + 2n_x$ with $j = 1, 2, \ldots, n_y$. All regions share the same transverse width $n_y$. Along the $x$ direction, the system is partitioned into three segments: the left reservoir, the central driven region, and the right reservoir. Here, $n_L$, $2n_x$, and $n_R$ denote the number of sites of the left reservoir, central region, and right reservoir along the $x$ direction, respectively, while $n_y$ denotes the transverse system size, which is taken to be the same for the system and the reservoirs.

    Each reservoir-$\lambda$ is described by the Hamiltonian
    \begin{equation}
    \label{app:hl}
    H_\lambda
    =
    \sum_{r, r'}
    \mathbf{a}^{\dagger}_{\lambda}(r)
    h_{\lambda}(r,r')
    \mathbf{a}_{\lambda}(r'),
    \end{equation}
    where $\mathbf{a}_{\lambda}(r)$ annihilates a fermion at site $r=(i,j)$ in reservoir $\lambda$. We assume that the hopping amplitudes along the $y$ direction vanish in $h_{\lambda}(r,r')$, so that each reservoir decomposes into $n_y$ independent one-dimensional noninteracting fermionic chains. The reservoirs are characterized by Fermi distribution functions
    \begin{align}
        f_\lambda(\omega)
        =
        \frac{1}{1+e^{(\omega - \mu_\lambda)/\mathcal{T}_\lambda}},
    \end{align}
    where $\mu_\lambda$ and $\mathcal{T}_\lambda$ denote the chemical potential and temperature of reservoir $\lambda$.

     The coupling between the system and the $\lambda$-th reservoir is described by
    \begin{equation}
    H_{S\lambda}
    =
    \sum_{\mathbf{r},\mathbf{r}'}
    c^\dagger(\mathbf{r})\,
    \mathcal{V}^{\lambda}_{\mathbf{r},\mathbf{r}'}
    \,a_{\lambda}(\mathbf{r}')
    + \mathrm{h.c.},
    \end{equation}
    where $c^\dagger(\mathbf r)$ creates a fermion at lattice site $\mathbf r=(x,y)$ in the system and implicitly carries the sublattice degree of freedom through the parity of $x+y$, i.e. the lattice sites satisfying $x+y$ even (odd) belong to sublattice $A$ ($B$).
    
  The tunneling matrices couple only boundary sites of the system to the reservoirs and, for the case of 1D reservoirs, are given by
    \begin{align}
    \label{eq:VL}\mathcal{V}^{L}_{\mathbf{r},\mathbf{r}'}
    &=
    t_L\,
    \delta_{\mathbf{r},(n_L+1,j)}\,
    \delta_{\mathbf{r}',(n_L,j)}, \\
    \mathcal{V}^{R}_{\mathbf{r},\mathbf{r}'}\label{eq:VR}
    &=
    t_R\,
    \delta_{\mathbf{r},(n_L+2n_x,j)}\,
    \delta_{\mathbf{r}',(n_L+2n_x+1,j)},
    \end{align}
    with $j=1,\ldots,n_y$. Here $t_{L(R)}$ denotes the tunneling amplitude to the left (right) reservoir. This choice of coupling connects the boundary sites of the system to the adjacent sites of the corresponding reservoirs.

    We now focus on the transport properties of the driven junction. To compute the current flowing between the reservoirs, we employ the Floquet NEGF formalism. The Heisenberg equations of motion for the system operators $\mathbf{c}(t)$ and reservoir operators $\mathbf{a}_{\lambda}(t)$ read
    \begin{eqnarray}\label{eq:eomsys1}
    i \dot{\mathbf{c}}(t)
    &=&
    h(t)\mathbf{c}(t) + \sum_\lambda \mathcal{V}^{\lambda}\mathbf{a}_{\lambda}(t),\\
    \label{eq:eomres1}
    i \dot{\mathbf{a}}_{\lambda}(t)
    &=&
    h_{\lambda}\mathbf{a}_{\lambda}(t)
    + \mathcal{V}^{\lambda\dagger}\mathbf{c}(t).
    \end{eqnarray}
    Here $\mathbf{c}(t)$ and $\mathbf{a}_{\lambda}(t)$ denote column vectors of annihilation operators in the system and reservoir $\lambda$, respectively. Solving Eq.~\eqref{eq:eomres1} yields
    \begin{align}\label{eq:eqm0}
    \mathbf{a}_{\lambda}(t)
    =
    i\, g^{\lambda}(t - t_0)\, \mathbf{a}_{\lambda}(t_0)
    + \int_{t_0}^{t} dt'\,
    g^{\lambda}(t - t')\, \mathcal{V}^{\lambda\dagger}\mathbf{c}(t'),
    \end{align}
    where $g^{\lambda}(t - t') = -i e^{-i(t - t') h_{\lambda}} \Theta(t - t')$ is the retarded Green's function of reservoir $\lambda$ and recall that $ h_{\lambda}$ appears in Eq.~\eqref{app:hl}. Taking $t_0 \to -\infty$ ensures that the reservoirs are coupled to the system in the distant past.\\
    
    Substituting Eq.~\eqref{eq:eqm0} into Eq.~\eqref{eq:eomsys1}, we obtain
    \begin{align}\nonumber
    \label{eq:sysresconn_a}
    [i\partial_t - h(t)]\, \mathbf{c}(t)
    + \sum_{\lambda}\int_{t_0}^{t} dt'\, \Sigma_\lambda(t - t')\, \mathbf{c}(t')\\
    =
    \sum_{\lambda} \mathcal{V}^{\lambda}\, \xi^{\lambda}(t),
    \end{align}
    where 
    \begin{equation}
    \xi^{\lambda}(t)= i g^{\lambda}(t - t_0)\mathbf{a}_{\lambda}(t_0)\label{eq:Xi_lambda}\,,
    \end{equation}
    and 
    \begin{align}
        \Sigma_{\lambda}(t - t') = 
\mathcal{V}^{\lambda} g^{\lambda}(t - t') \mathcal{V}^{\lambda\dagger}
\label{app:gr}
    \end{align}
    is the self-energy of the reservoirs. In the Fourier space Eq.~\eqref{app:gr} can be written as: 
   \begin{align}
        \Sigma_{\lambda}(\omega) = 
        \mathcal{V}^{\lambda} g^{\lambda}(\omega) \mathcal{V}^{\lambda\dagger}.\label{eq:Sigma_FS}
   \end{align}
   We consider the reservoirs in the wide-band limit, in which Green's function of the reservoirs [that appears in Eq.~\eqref{eq:Sigma_FS}] in the frequency space is approximated as $g^{\lambda}(\omega) = -i\pi\rho^{\lambda}$, where $\rho^{\lambda}$ is the constant density of states of reservoir $\lambda$.
  This yields a frequency-independent self-energy term in the weak coupling limit, i.e. Eq.~\eqref{eq:Sigma_FS} becomes
  \begin{align}\label{eq:Sigma_FI}
       \Sigma_{\lambda} &= -i\pi \mathcal{V}^{\lambda} \rho^{\lambda} \mathcal{V}^{\lambda\dagger}, 
  \end{align}
  Furthermore, the above matrix equation can be explicitly written (using equations~\eqref{eq:VL}-\eqref{eq:VR}) for $\lambda=L, R$, as 
    \begin{align}\label{eq:Sigma_L}
      \Sigma_{L,xx',yy'} &= -i\pi t_L^2 \rho^{L}\delta_{xx'}\delta_{yy'}\delta_{x,n_L+1}\\\label{eq:Sigma_R}
      \Sigma_{R,xx',yy'} &= -i\pi t_R^2 \rho^{R}\delta_{xx'}\delta_{yy'}\delta_{x,n_L+2n_x+1}
    \end{align}
    Using equations~\eqref{eq:Sigma_L}-\eqref{eq:Sigma_R}, the $\Gamma_\lambda$ matrices, used in the conductance expression given in the Eq.~\eqref{eq:G-transmission-tro-terminal} of the main text,
    can be expressed as 
    \begin{align}
    \label{eq:Gamma_Lambda}\Gamma_{\lambda}=\frac{\Sigma_{\lambda}-\Sigma^{\dagger}_{\lambda}}{2\pi i}.
    \end{align}
    Eq.~\eqref{eq:Gamma_Lambda} can be also written component wise, for $\lambda=L, R$ as,   
    \begin{align}
        \Gamma_{L,xy,x'y'}&=\gamma_L \delta_{xx'}\delta_{yy'}\delta_{x,1},\\
        \Gamma_{R,xy,x'y'}&=\gamma_R \delta_{xx'}\delta_{yy'}\delta_{x,2 n_x},
    \end{align}
     with $\gamma_{\lambda}=t_{\lambda}^2\rho^{\lambda}$ as the single system-bath coupling parameter. For our numerics we have taken $t_L=t_R$ and also $\rho^L=\rho^R$, i.e. $\gamma_L=\gamma_R=\gamma$.

    In the wide band limit, the self energy term of the reservoirs becomes frequency independent [see Eq.~\eqref{eq:Sigma_FI}], and therefore Eq.~\eqref{eq:sysresconn_a} reduces to
\begin{equation}\label{eq:sysresconn1a}
    [i\partial_t - h(t) + i\Gamma]\, \mathbf{c}(t)
    =
    \sum_{\lambda} \mathcal{V}^{\lambda} \xi^{\lambda}(t),
    \end{equation}
    where $\Gamma=\Gamma_L+\Gamma_R$ and we recall that $\xi^{\lambda}(t)$ is given in Eq.~\eqref{eq:Xi_lambda}. The solution for the system operators can be expressed in terms of the retarded Green's function as
    \begin{align}
    \mathbf{c}(t)
    =
    \sum_{\lambda}
    \int_{t_0}^{t} dt'\,
    \mathbf{G}(t,t')\,
    \mathcal{V}^{\lambda} \xi^{\lambda}(t').\label{eq:ct}
    \end{align}
    where the retarded Green's function $\mathbf{G}(t,t')$ in Eq.~\eqref{eq:ct} satisfies
    \begin{align}
    [i\partial_t - h(t) + i\Gamma]\,
    \mathbf{G}(t,t') = \delta(t - t').\label{eq:G_def}
    \end{align}
    To elaborate Eqs.~\eqref{eq:sysresconn1a}-\eqref{eq:G_def} further, let us define the operator $L_t = i\partial_t - h(t) + i\Gamma$. Applying this operator on both sides of Eq.~\eqref{eq:ct} gives 
    \begin{align}
    \label{eq:app_LG}
    L_t \mathbf{c}(t)=\sum_{\lambda}
    \int_{t_0}^{t} dt'\,
    L_t\mathbf{G}(t,t')\,
    \mathcal{V}^{\lambda} \xi^{\lambda}(t')
    \end{align}
    Now, using the definition of the Green's function in Eq.~\eqref{eq:G_def}, i.e. $L_t G(t,t')=\delta(t-t')$ in Eq.~\eqref{eq:app_LG} we get
    \begin{align}
    L_t \mathbf{c}(t)=\sum_{\lambda}
    \int_{t_0}^{t} dt'\,
    \delta(t-t')\,
    \mathcal{V}^{\lambda} \xi^{\lambda}(t')=\sum_{\lambda}
    \mathcal{V}^{\lambda} \xi^{\lambda}(t),
    \end{align}
    which is Eq.~\eqref{eq:ct}.

    The Floquet components $\mathbf{G}^{(q)}(\omega)$ are defined via the two-time Fourier expansion
    \begin{align}
        \mathbf{G}(t,t')
    &=
    \int_{-\infty}^{\infty}\frac{d\omega}{2\pi}\;
    \mathbf{G}(t,\omega)
    e^{-i\omega(t-t')},\label{eq:Gttp-1}\\
    &=
    \sum_{q\in\mathbb{Z}}
    \int_{-\infty}^{\infty}\frac{d\omega}{2\pi}\;
    \mathbf{G}^{(q)}(\omega)
    e^{-i\omega(t-t')}e^{-iq\Omega t},\label{eq:Gttp}
    \end{align}
    where we recall that $\Omega=2\pi/T$ is the frequency of the drive.
    
    The Green's function can be written in the spectral form~\cite{Kohler2005}
    \begin{align}\label{eq:Green_fourier}
    \mathbf{G}^{(q)}(\omega)
    =
    \sum_{\alpha,n\in\mathbb{Z}}
    \frac{|u_{\alpha}^{-\,(n+q)}\rangle
    \langle u_{\alpha}^{+(n)}|}
    {\omega - e^{(n)}+ i \gamma_{\alpha}},
    \end{align}
    where $e_{\alpha}^{(n)}=e_{\alpha}+n\Omega$. The Floquet modes $|u_{\alpha}^{\pm}(t)\rangle$ satisfy
    \begin{eqnarray}
    \left(h(t) - i\Gamma - i\partial_t \right) |u^{-}_\alpha(t)\rangle
    &=&
    (e_\alpha - i\gamma_\alpha) |u^{-}_\alpha(t)\rangle,\\
    \left(h(t) + i\Gamma - i\partial_t \right)|u_\alpha^{+}(t)\rangle
    &=&
    (e_\alpha + i\gamma_\alpha) |u_\alpha^{+}(t)\rangle,
    \end{eqnarray}
    and form a biorthonormal complete basis,
    \begin{equation}
    \langle u_{\beta}^{+}(t)|u^{-}_{\alpha}(t)\rangle=\delta_{\alpha\beta},
    \qquad
    \sum_{\alpha}|u^{+}_{\alpha}(t)\rangle\langle u_{\alpha}^{-}(t)|
    =\mathbf{I}.
    \end{equation}

	\section{Bond currents using NEGF}\label{sec:bond_currents_using_negf}
    In this section, we derive the bond-resolved currents for a periodically driven system coupled to static reservoirs. These results form the technical foundation for the bond-resolved current expressions in Eqs.~\eqref{eq:bond-current-xy-a}--\eqref{eq:bond-current-xy-b} of the main text. Using these expressions, we compute the two-terminal (longitudinal) and Hall (transverse) bond conductances, which are discussed in Sec.~\ref{sec:Numerical results} and illustrated in Fig.~\ref{fig:FIG_G2GH_bulk}.

    In addition to the current at the interfaces between the system and the reservoirs, one can compute bond currents within the driven central region. For a time-periodic Hamiltonian $H(t+T)=H(t)$, the local charge density operator at each lattice site $\mathbf r=(x,y)$ is given by
    $\mathcal{N}(\mathbf r,t) = c^\dagger(\mathbf r,t)c(\mathbf r,t)$,
    where $x=n_L+1,\ldots,n_L+2n_x$ and $y=1,\ldots,n_y$. We recall that the lattice possesses a bipartite ($A/B$) sublattice structure, which is implicitly encoded in the parity of $x+y$. Using the Heisenberg equation $\dot{\mathcal{N}}(\mathbf r,t)=i[H(t),\mathcal{N}(\mathbf r,t)]$, together with the standard tight-binding continuity equation (with $e=\hbar=1$), the bond current from $\mathbf r$ to $\mathbf r'$ is given by
    \begin{equation}
    J_{\mathbf r\to \mathbf r'}(t)
    =
    2\,\mathrm{Im}\!\left[
    H_{\mathbf r,\mathbf r'}(t)\,
    \langle c^\dagger(\mathbf r,t)c(\mathbf r',t)\rangle
    \right],\label{eq:J_e1}
    \end{equation}
    where $H_{\mathbf r,\mathbf r'}(t)$ denotes the single-particle matrix element of the system Hamiltonian between sites $\mathbf r$ and $\mathbf r'$. The time-averaged bond current using Eq.~\eqref{eq:J_e1} is then
    \begin{equation}
    \mathcal{J}_{\mathbf r\to \mathbf r'}
    =
    \frac{1}{T}\int_0^T dt\,
    2\, \mathrm{Im}\!\left[
    H_{\mathbf r,\mathbf r'}(t)
    \langle c^\dagger(\mathbf r,t)c(\mathbf r',t)\rangle
    \right] .\label{eq:Javg}
    \end{equation}
    Using Eq.~\eqref{eq:ct} and the Floquet Green's function given in Eq.~\eqref{eq:Gttp-1}, the equal-time correlation function that appears in Eq.~\eqref{eq:Javg} can be written as
    \begin{equation}
    \langle c^\dagger(\mathbf r,t)c(\mathbf r',t)\rangle
    =
    \sum_{\lambda}\int_{-\infty}^{\infty} \frac{d\omega}{2\pi}\;
    \chi(t,\omega)_{\lambda,\mathbf r',\mathbf r}f_\lambda(\omega),
    \end{equation}
    where
    \begin{align}
    \chi(t,\omega)_{\lambda,\mathbf r',\mathbf r}
    =
    2\pi \Big[
    G(t,\omega)\,\Gamma_\lambda\,G^\dagger(t,\omega)\Big]_{\mathbf r', \mathbf r}\,
    \end{align}
    is the two point correlation function. 
    We now consider the linear-response regime, where the left and right reservoirs are maintained at chemical potentials
    \(
    \mu_L=\mu_F+\Delta\mu
    \)
    and
    \(
    \mu_R=\mu_F
    \),
    respectively, with
    \(
    \Delta\mu\rightarrow 0
    \).
    The current can then be decomposed into equilibrium and nonequilibrium contributions,
    \begin{align}
    	\mathcal{J}_{\mathbf r\to \mathbf r'}
    	=
    	\mathcal{J}^{\mathrm{eq}}_{\mathbf r\to \mathbf r'}
    	+
    	\mathcal{J}^{\mathrm{ex}}_{\mathbf r\to \mathbf r'}.
    	\label{eq:Jdecomp_appendix}
    \end{align}
    At equilibrium,
    \(
    f_L(\omega)=f_R(\omega)=f_F(\omega)
    \),
    where
    \(
    f_F(\omega)
    =
    [\exp((\omega-\mu_F)/T)+1]^{-1}
    \)
    is the Fermi-Dirac distribution function. The corresponding equilibrium contribution is
    \begin{align}
    	\mathcal{J}^{\mathrm{eq}}_{\mathbf r\to \mathbf r'}
    	=
    	\frac{1}{T}
    	\int_0^T dt
    	\int_{-\infty}^{\infty} d\omega\;
    	2\,\mathrm{Im}
    	\Big[
    	H_{\mathbf r,\mathbf r'}(t)\,
    	\chi^{(0)}(t,\omega)_{\mathbf r',\mathbf r}
    	\Big],
    	\label{eq:Jeq_appendix}
    \end{align}
    where $\chi^{(0)}(t,\omega)=2\pi \left[G(t,\omega)(\Gamma_L+\Gamma_R)G^\dagger(t,\omega)\right]f_F(\omega)$, which does not contribute to the net current due to equilibrium symmetry. The nonequilibrium/excess contribution arises entirely from the $\Delta\mu$-bias between the reservoirs, i.e. \(f_L(\omega+\Delta\mu_F)=f_L(\omega)-\Delta\mu\,\partial_\omega f_L(\omega)|_{\mu_F}\), and can be written as
    \begin{align}
    	\frac{\mathcal{J}^{\mathrm{ex}}_{\mathbf r\to \mathbf r'}}{\Delta\mu}
    	\!=\!&-\!
    	\int_0^T \frac{dt}{T}
    	\int_{-\infty}^{\infty} \!\!d\omega\;
    	2\,\mathrm{Im}
    	[H_{\mathbf r,\mathbf r'}(t)
    	\Lambda(t,\omega)_{\mathbf r',\mathbf r}]
    	\partial_\omega f_L(\omega)
    	\label{eq:Jex_appendix}
    \end{align}
    where $\Lambda(t,\omega)= 2\pi G(t,\omega)\,\Gamma_L\,G^\dagger(t,\omega)$. At zero temperature,
    \(
    -\partial_\omega f_F(\omega)
    =
    \delta(\omega-\mu_F)
    \). Substituting Eq.~\eqref{eq:Jex_appendix}, in Eq.~\eqref{eq:Jdecomp_appendix}, we obtain the linear-response expression
    \begin{equation}
        \mathcal{J}_{\mathbf r\to \mathbf r'}(\mu_F)
        =
        \frac{1}{T}\int_0^T dt\;
        2\,\mathrm{Im}\Big[
        H_{\mathbf r,\mathbf r'}(t)\,
        \Lambda(L,t,\mu_F)_{\mathbf r',\mathbf r}
        \Big]\Delta\mu.\label{eq:Jex2}
    \end{equation}
    
    We now expand the time-dependent terms that appears in the integrand of Eq.~\eqref{eq:Jex2}, in the Fourier space. Note that 
    \begin{align}
    H(t)&=\sum_{p}e^{-ip\Omega t}H^{(p)},\label{eq:hft}\\
    G(t,\omega)&=\sum_{k}e^{-ik\Omega t}G^{(k)}(\omega),\label{eq:gft}
    \end{align}
    with $H^{(p)}=\frac{1}{T}\int_0^T dt\,e^{ip\Omega t}H(t)$. Using Eq.~\eqref{eq:hft} and Eq.~\eqref{eq:gft} in Eq.~\eqref{eq:Jex2}, and after integrating it over time we obtain 
    \begin{align}\nonumber
    \mathcal{J}_{\mathbf r\to \mathbf r'}(\mu_F)
    =&
    \sum_{k,l}
    2\,\mathrm{Im}\Big(
    H^{(l-k)}_{\mathbf r,\mathbf r'}\,
    \big[G^{(k)}(\mu_F)\Gamma_L \\
    &G^{\dagger(l)}(\mu_F)\big]_{\mathbf r',\mathbf r}
    \Big)\Delta\mu.\label{eq:Jbond_ft}
    \end{align}
    Using the spectral representation of the Floquet Green's function given in Eq.~\eqref{eq:Green_fourier}, the bond current in Eq.~\eqref{eq:Jbond_ft} can be written as
    \begin{align}\nonumber
    \mathcal{J}_{\mathbf r\to \mathbf r'}(\mu_F)
    =&
    \sum_{k,l,n,m,\alpha,\beta}
    2\,\mathrm{Im}\Bigg[
    \langle \mathbf r|u_\beta^{-\,(l+n)}\rangle
    \langle u_\beta^{-\,(l+n)}|H^{(l-k)}\\
    &|\mathbf r'\rangle \times
    \langle \mathbf r'|u_\alpha^{+\,(k+m)}\rangle
    \frac{\langle u_\alpha^{+(m)}|\Gamma_L|u_\beta^{+(n)}\rangle}
    {Z_{\alpha,+}^{(m)}\,Z_{\beta,-}^{(n)}}
    \Bigg]\Delta\mu.\label{eq:Jexp}
    \end{align}
    Here $Z_{\alpha,+}^{(m)}=(\mu_F-e_\alpha^{(m)}+i\gamma_\alpha)$ and
    $Z_{\beta,-}^{(n)}=(\mu_F-e_\beta^{(n)}-i\gamma_\beta)$. The quantities, such as 
    $\langle \mathbf r | u_\alpha^{+\,(n)} \rangle$ and $\langle \mathbf r | u_\alpha^{-\,(n)} \rangle$, that appear in Eq.~\eqref{eq:Jexp}, denotes the projection of the $n$-th Floquet harmonic of the left and right Floquet modes onto lattice site $\mathbf r$.
    After relabeling indices and regrouping terms in Eq.~\eqref{eq:Jexp}, the current can be expressed in the compact form as
    \begin{equation}
    \mathcal{J}_{\mathbf r\to \mathbf r'}(\mu_F)
    =
    \sum_{q,\alpha,\beta}
    2\,\mathrm{Im}\Big[
    H^{(q)}_{\mathbf r,\mathbf r';\beta\alpha}
    \mathcal{F}^{(q)}_{\alpha\beta}(\mu_F)
    \Big]\Delta\mu,\label{eq:Jcomact}
    \end{equation}
    where
    \begin{align}
    \mathcal{F}^{(q)}_{\alpha\beta}(\mu_F)
    &=
    \sum_{n}
    \frac{\langle u_\alpha^{+(n+q)}|\Gamma_L|u_\beta^{+(n)}\rangle}
    {Z_{\alpha,+}^{(n+q)}\,Z_{\beta,-}^{(n)}},\\[4pt]
    H^{(q)}_{\mathbf r,\mathbf r';\beta\alpha}
    &=
    \sum_{k,l}
    \langle u_\beta^{-\,(l)}|\mathbf r\rangle
    \langle \mathbf r|H^{(l-k+q)}|\mathbf r'\rangle
    \langle \mathbf r'|u_\alpha^{-\,(k)}\rangle.
    \end{align}
    
    The longitudinal and transverse conductances in the linear-response regime are defined in units of $e^2/h$ as 
    \begin{align}
    \mathcal G_{x(y)}(\mu_F) = \frac{2\pi \mathcal J_{x(y)}(\mu_F)}{\Delta\mu},\label{eq:Gbond}
    \end{align}
    where $\mathcal J_{x}$ and $\mathcal J_y$, can be obtained from Eq.~\eqref{eq:Jcomact}, by setting $(\mathbf r, \mathbf r')$ to be $((x,y), (x+1,y))$ and $((x,y), (x,y+1))$, respectively. We can write the bond conductances  given in Eq.~\eqref{eq:Gbond} as
    \begin{align}
    \mathcal{G}_x(\mu_F) &= \sum_{y=1}^{n_y}\sum_{q,\alpha,\beta}
    4\pi\,\mathrm{Im}\Big[
    H^{(q)}_{(x,y),(x+1,y);\beta\alpha}
    \mathcal{F}^{(q)}_{\alpha\beta}(\mu_F)
    \Big],\\
    \mathcal{G}_y(\mu_F) &= \sum_{x=n_L+1}^{n_L+n_x}\sum_{q,\alpha,\beta}
    4\pi\,\mathrm{Im}\Big[
    H^{(q)}_{(x,y),(x,y+1);\beta\alpha}
    \mathcal{F}^{(q)}_{\alpha\beta}(\mu_F)
    \Big].
    \end{align}
    We would like to emphasis here, that the above discussion was for the strip geometry shown in Fig.~\ref{fig:FIG-SETUP}.

    Similarly, for a cylindrical geometry with periodic boundary conditions along the $y$ direction [the corresponding system setup is shown in Fig.~\ref{fig:cylindrical_geo}], the bond conductance expression can be obtained following the same procedure. The longitudinal conductance can be written as
    \begin{equation}
    \mathcal{G}_x(\mu_F)
    =
    \sum_{k_y,q,\alpha,\beta}
    4\pi\,\mathrm{Im}\!\Big[
    H^{(q)}_{x,x+1;\alpha\beta}(k_y)\,
    \mathcal{F}^{(q)}_{\beta\alpha}(k_y;\mu_F)
    \Big],
    \label{eq:Jx_pbc}
    \end{equation}
    and the transverse conductance averaged over $y$ is defined as $\mathcal{G}^{\mathrm{avg}}_y(\mu_F)=\frac{1}{n_y}\sum_{y=1}^{n_y}\mathcal{G}_y(\mu_F)$, i.e.
    \begin{equation}
    \mathcal{G}^{\mathrm{avg}}_y(\mu_F)
    =\frac{2\pi}{n_y}
    \sum_{x,k_y,q,\alpha,\beta}
    2\,\mathrm{Im}\!\Big[
    \partial_{k_y} H^{(q)}_{x,x;\alpha\beta}(k_y)\,
    \mathcal{F}^{(q)}_{\beta\alpha}(k_y;\mu_F)
    \Big].
    \label{eq:Jy_pbc}
    \end{equation}
    Here, $H^{(q)}(k_y)=\frac{1}{T}\int_0^T dt\, e^{iq\Omega t} H(t,k_y)$ denotes the $q$-th Floquet component of the Hamiltonian in cylinder geometry, given in Eq.~\eqref{eq:h5_kxkyt_cyl}. Also, the coupling matrices in this setup for $\lambda=L, R$, are given by following    
    \begin{align}
        \Gamma_{L,x,x'}&=\gamma_L \delta_{xx'}\delta_{x,1},\\
        \Gamma_{R,x,x'}&=\gamma_R \delta_{xx'}\delta_{x,2 n_x}.
    \end{align}
    \newpage
	\bibliography{arxiv.bib}	
\end{document}